# Development of a Selective Wet-Chemical Etchant for 3D Structuring of Silicon via Nonlinear Laser Lithography

Mona Zolfaghari Borra[1,2]
[1]ODTÜ-GÜNAM, Middle East Technical University, 06800, Ankara, Turkey
[2]Micro and Nanotechnology Graduate Program of Natural and Applied Sciences, Middle East Technical University (METU), 06800, Ankara, Turkey
Email: mborra@metu.edu.tr

Behrad Radfar[1,2,3]
[1]ODTÜ-GÜNAM, Middle East Technical University, 06800, Ankara, Turkey
[2]Micro and Nanotechnology Graduate Program of Natural and Applied Sciences, Middle East Technical University (METU), 06800, Ankara, Turkey
[3]Department of Electronics and Nanoengineering, Aalto University, Tietotie 3, FI-02150 Espoo, Finland
Email: behrad.radfar@aalto.fi

Hisham Nasser[1]
[1]ODTÜ-GÜNAM, Middle East Technical University, 06800, Ankara, Turkey
Email: nasser@metu.edu.tr

Tahir Çolakoğlu[1,4]
[1]ODTÜ-GÜNAM, Middle East Technical University, 06800, Ankara, Turkey
[4]Department of Physics Engineering, Ankara University, 06100, Ankara, Turkey
Email: tcolakoglu@ankara.edu.tr

Onur Tokel[5,6]
[5]Department of Physics, Bilkent University, 06800, Ankara, Turkey
[6]UNAM National Nanotechnology Research Center and Institute of Materials Science and Nanotechnology, Bilkent University, Ankara 06800, Turkey
Email: otokel@bilkent.edu.tr

Ahmet Turnalı[7,8]
[7]Department of Electrical and Electronics Engineering, Bilkent University, 06800, Ankara, Turkey
[8]Department of Electrical and Computer Engineering, Boston University, 8 Saint Mary's Street, Boston, MA 02215, USA
Email: turnali@bu.edu

Merve Demirtaş[9,10]
[9]Department of Physics, Middle East Technical University (METU), 06800, Ankara, Turkey
[10]Network Technologies Department, TUBITAK ULAKBIM, 06800, Ankara, Turkey
Email: merve.demirtas@tubitak.gov.tr

Hande Ustunel[2,9]
[2]Micro and Nanotechnology Graduate Program of Natural and Applied Sciences, Middle East Technical University (METU), 06800, Ankara, Turkey
[9]Department of Physics, Middle East Technical University (METU), 06800, Ankara, Turkey
Email: ustunel@metu.edu.tr

Daniele Toffoli[11,12]
[11]Department of Chemical and Pharmaceutical Sciences, Trieste University, Trieste, Italy
[12]IOM-CNR, Istituto Officina dei Materiali, 34149 Trieste, Italy

Email: toffoli@units.it

F. Ömer İlday[5,7]
[5]Department of Physics, Bilkent University, 06800, Ankara, Turkey
[7]Department of Electrical and Electronics Engineering, Bilkent University, 06800, Ankara, Turkey
Email: ilday@bilkent.edu.tr

Raşit Turan[1,2,9]
[1]ODTÜ-GÜNAM, Middle East Technical University, 06800, Ankara, Turkey
[2]Micro and Nanotechnology Graduate Program of Natural and Applied Sciences, Middle East Technical University (METU), 06800, Ankara, Turkey
[9]Department of Physics, Middle East Technical University (METU), 06800, Ankara, Turkey
Email: turanr@metu.edu.tr

Ihor Pavlov[1,2,9]
[1]ODTÜ-GÜNAM, Middle East Technical University, 06800, Ankara, Turkey
[2]Micro and Nanotechnology Graduate Program of Natural and Applied Sciences, Middle East Technical University (METU), 06800, Ankara, Turkey
[9]Department of Physics, Middle East Technical University (METU), 06800, Ankara, Turkey
Email: ipavlov@metu.edu.tr

Alpan Bek[1,2,9]
[1]ODTÜ-GÜNAM, Middle East Technical University, 06800, Ankara, Turkey
[2]Micro and Nanotechnology Graduate Program of Natural and Applied Sciences, Middle East Technical
[9]Department of Physics, Middle East Technical University (METU), 06800, Ankara, Turkey
Email: bek@metu.edu.tr

## Abstract

Recently-demonstrated high-quality three-dimensional (3D) subsurface laser processing inside crystalline silicon (c-Si) wafers opens a door to a wide range of novel applications in multidisciplinary research areas. Using this technique, a novel maskless micro-pillars with precise control on the surface reflection and coverage are successfully fabricated by etching the laser processed region of c-Si wafer. To achieve this, a particular selective wet chemical etching is developed to follow subsurface laser processing of c-Si to reveal the desired 3D structures with smooth surfaces. Here, we report the development of a novel chromium-free chemical etching recipe based on copper nitrate, which yields substantially smooth surfaces at high etch rate and selectivity on the both laser-processed Si surface and subsurface, i.e., without significant etching of the unmodified Si. Our results show that the etch rate and surface morphology are interrelated and strongly influenced by the composition of the adopted etching solution. After an extensive compositional study performed at room temperature, we identify an etchant with a selectivity of over 1600 times for laser-modified Si with respect to unmodified Si. We also support our findings using density functional theory calculations of HF and Cu adsorption energies, indicating significant diversity on the c-Si and laser-modified surfaces.

## 1. Introduction

Recently, we have succeeded directly to laser-write 3D microstructures deep inside Si through a method that relies on a self-organized mechanism [1], [2] while the efforts dating until 2005 [3] were largely unsuccessful. Our method is an adaptation of another self-organized technique, known as nonlinear laser lithography that we developed to create self-organized sub-wavelength structures on surfaces [4]. In both 2D and 3D implementations of nonlinear laser lithography, we exploit an intrinsic feedback mechanism, where the pulses cause changes in the material that alters how the subsequent pulses focus. In the case of 3D Si modification, these changes constitute rod-like elongated structures along the beam propagation direction. The length of the rods can be changed by the laser parameters and the number of incident pulses. They can also be positioned at an arbitrary depth below the surface where the laser modification results in a change of the materials optical index. Thus, we have demonstrated a variety of structures of interest for various applications, including optical waveguides, information storage, holographic structures, Fresnel lenses for the near-IR, among others. We have further shown that the laser-modified regions of Si can be selectively etched to reveal various 3D structures [1]. Today, laser-processing of Si is an active area where a wide variety of applications have been demonstrated and reviewed in [5], including quantum photonics [6], lab-on-chip [7]–[10], optical communications [11], and microfluidics [12]–[14]. An emerging application demonstrated by this technique is laser-assisted Si slicing (LASIS) [1], in which the fabricated thin c-Si slices can be processed into thin/ultra-thin Si solar cells produced with a minimum loss of Si material [15], [16]. Low material loss in slicing is an important requirement in reducing the solar cell cost. However, the vast majority of the studies on laser-based 3D processing of Si are limited to the creation of optical index variations with few demonstrations of the chemical etching to create standalone 3D structures. For revealing the 3D laser sculpted structures out of Si bulk, laser-modified regions of the c-Si subsurface are removed by selective chemical etching. We attribute the slow development of this area to the lack of a well-developed etchant, which we address here. In contrast to 3D or volume laser processing, surface laser processing is a well-established technique to create desired features on the Si surface. For instance, laser-induced periodic surface structuring (LIPSS) is used to create light trapping interfaces for solar cells [17]–[21] or fabrication black-Si [22]–[24].

In laser assisted processing of Si, ultra-short near to mid-infrared laser pulses are focused into the bulk of the wafer. The interaction of the intense localized laser radiation with Si induces non-linear effects such as multi-photon absorption, self-focusing, and plasma generation inside the focal volume [25]–[27]. The subsurface laser processing provides precise local modifications at the targeted sites regardless of Si crystallinity or orientation [28], [29]. In this technique, the optimization of the laser parameters (i.e., laser wavelength, pulse duration, and pulse energy) enables precise control over the laser-Si interaction. The formation of subsurface structures using a nanosecond or femtosecond pulsed laser beam [30]–[34] is a new micro and nanofabrication approach [35], [36]for various applications which require high-quality structures. This method creates modification at localized regions with minor damage on the c-Si surface [1], [2].

So far, neither the surface nor subsurface laser processing of Si has yet been investigated enough to unearth the full potential of 3D laser processing for the micro-fabrication of Si devices. We lack a well-established method to remove the laser-modified regions without damaging unprocessed areas. A natural outcome of surface laser processing is the intentional damage it induces to some surface areas. Therefore, a specially tailored selective etchant must be developed to remove only the laser-damaged regions. Researchers have invested substantial efforts in Si technology to control the defect densities associated with re-growth and Si wafer processing [37]. Selective chemical etching is a simple route to reveal the number of defects in c-Si. Chemical treatments typically comprise a redox reaction to etch the defects selectively. In a more general view, it is of great essence that such a special etchant is developed to selectively remove laser-modified Si out of c-Si wafer to sculpt functional 3D structures.

Etching is an essential step in the micro- and nanofabrication of c-Si-based devices. Among different etching techniques, solution-based chemical etching offers a low-cost and facile alternative compared to conventional plasma-enhanced gas-phase etching counterparts while yielding a high surface and bulk material quality due to the absence of plasma damage. Achieving a Si surface as defect-free as possible at microscale is crucial to ensure high performance of Si devices, such as micro-electromechanical systems (MEMS) [38]–[41], micro-optics [42]–[44], photonic devices, integrated circuits, and photovoltaic solar cells[21], [45]–[51].

Similar to surface laser processing, a subsequent chemical etch is needed to reveal the subsurface structures without altering unprocessed Si regions for the aforementioned wide applications. To this end, we need an etchant with excellent selectivity (the ratio of the removal rate of laser processed and pristine regions) and a high etch rate. Thus far, typical defect etchant recipes comprise nitric acid ($HNO_3$) or chromium (Cr)-based chemical solutions, which both serve as the oxidizing agent of Si surface. The $HNO_3$-based etchants, including HNA [52] and Dash etch (1956) [53] are crystallographic orientation and dopant type independent; however, they require highly doped Si (HNA) or high-temperature and longer process duration (Dash). Other etchants use a Cr-based oxidizing agent to delineate defects and dislocations in Si, including the Sirtl etch (1961) [54] and Seiter etch (1977) [55]. While the Sirtl etch works only for <111> oriented surfaces, the Seiter etch works well on <100>.

A highly selective etching solution of Si defects in the advanced electronics industry is the MEMC etching (Monsanto Electronic Materials Company - 1987) [56] which is a copper nitrate – $Cu(NO_3)_2$-based etchant and applies to both p- and n-doped Si. MEMC etch has a long lifetime during room-temperature processes and generates undesired pits on both <111> and <100> oriented surfaces. Its planar etch rate is the same for all crystallographic directions, while the etching uniformity is similar to that of Sirtl. Moreover, as it is a Cr-free solvent, it is relatively environmentally friendly compared to Cr-based etchants. It is commonly used in the destructive characterization of Si-wafer quality in industrial applications. In MEMC etch, the $HNO_3$ oxidizing agent oxidizes the Si surface, and hydrofluoric acid (HF) dissolves the formed $SiO_2$ while adding $Cu(NO_3)_2$, in particular, assists in reducing the resultant point defects. Therefore, the amount of $Cu(NO_3)_2$ in the mixture dramatically influences the localized differential oxidation at the defect sites. Although a detailed explanation of the mechanism is unknown, it is commonly accepted that $CH_3COOH$ of MEMC acts as a moderation agent of the surface reaction through limiting the oxidation. When the etchant solution contains too much $CH_3COOH$, the concentration of the oxidizing agent decreases, resulting in the impairment of the oxidization. This precludes a high etch rate. If the amount of $CH_3COOH$ is too small, the etching becomes too fast, increasing the surface roughness [57].

We adopted the MEMC etch to develop an optimized etchant for 3D nonlinear laser lithography of Si. This solution is highly defect-sensitive, which increases its selectivity in favor of the laser-modified regions. The solution depends on the crystal orientation and the dopant type. The etch durations are reasonably short. The solution has a long lifetime, making refreshing the chemical concentration (known as spiking) unnecessary. It is also relatively less toxic and more environmentally friendly owing to being Cr-free. In contrast, the Sirtl and Seiter etches yield toxic, even carcinogenic reaction byproducts [58]. Their use is heavily regulated in many countries.

This work focuses on the development of a chemical solution for selective etching of the laser-modified surface and subsurface Si regions at high etch rates. We first present an ultimate potential of the proposed development in 3D sculpting using our champion etchant. Our champion etchant is an optimized etching recipe which meets the requirements of reasonable etch rate and smooth Si facets. We support our findings via density functional theory (DFT) calculations that provide an atomistic-level understanding on how the etchant components interact with the surfaces in question.

## 2. Prospective of 3D sculpting

Micro-pillar arrays have been of interest for solar applications due to the increased light trapping and high carrier collecting efficiency inherent in the pillar structure. Light trapping of the incoming solar light can be achieved by texturing the front surface of the solar cell. Numerous methods are available, including nanowires [59], inverted pyramids [60], plasmonic metal NPs [47], etc. As an application of 3D sculpting, we demonstrate the laser-based fabrication of micron-size pillars array on c-Si wafers. Micro-pillar arrays on the surface of Si wafers can be used in solar cell design in which light can be effectively trapped via multiple reflections between pillars. Another application can be the fabrication of MEMS resonators for various applications requiring mechanical modulation [61], [62]. This technology allows obtaining a controllable size/depth combination of the pillar in the various ranges in two-step fabrication in Si substrate. Figure 1 depicts the fabrication procedure of micro-pillar arrays. The first step is laser processing which induces structural modification of Si. Negative pillar structures are directly written by a raster scan of the focal spot on and in the Si wafer (Figure 1 (a)). Then, in the second step, removal of the laser written regions by our developed selective etchant is used to extract the Si micro-pillar structure (Figure 1 (b)). For solar cell applications, the light trapping characteristics of the micro-pillar array can be investigated against varying depth and size of micro-pillar arrays (Figure 2). With this technique, micro-pillars with an area as small as ~$9 \times 9$ $\mu m^2$ and with a depth as large as 118 $\mu m$ were successfully fabricated yielding an impressive aspect ratio >13.

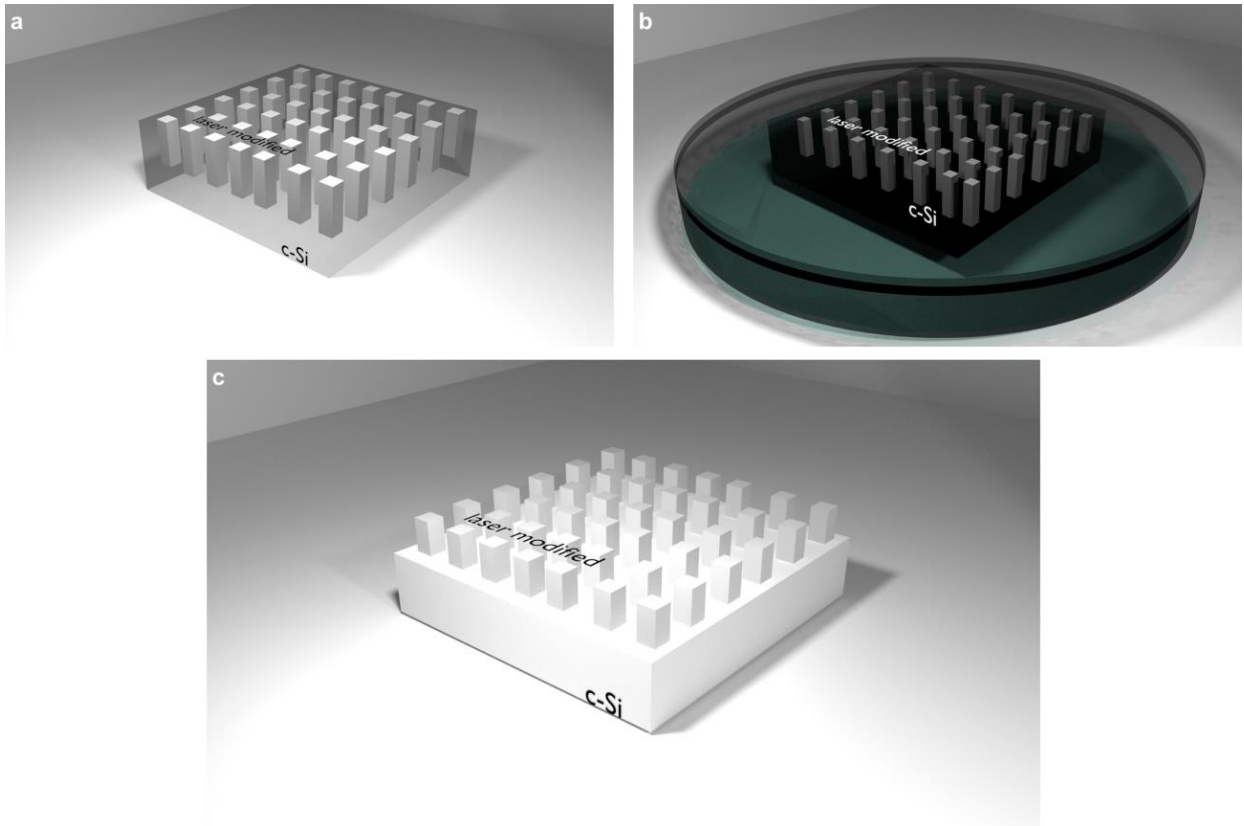

**Figure 1. Steps of micro-pillar fabrication: (a) Modified part after laser processing pattern of pillar; (b) Selective chemical etching; (c) Pillar arrays.**

Figure 2 shows micro-pillar arrays of 9 µm × 9 µm, 31 µm × 31 µm, and 57 µm × 57 µm pillar area. Since density of pillars (total number of pillars in designed area) is constant for all pillar sizes, surface coverages of ~5%, ~25% and ~56.25% are achieved, respectively. 2 mm × 2 mm micro-pillar arrays are revealed in 20 minutes of etch duration for all sizes.

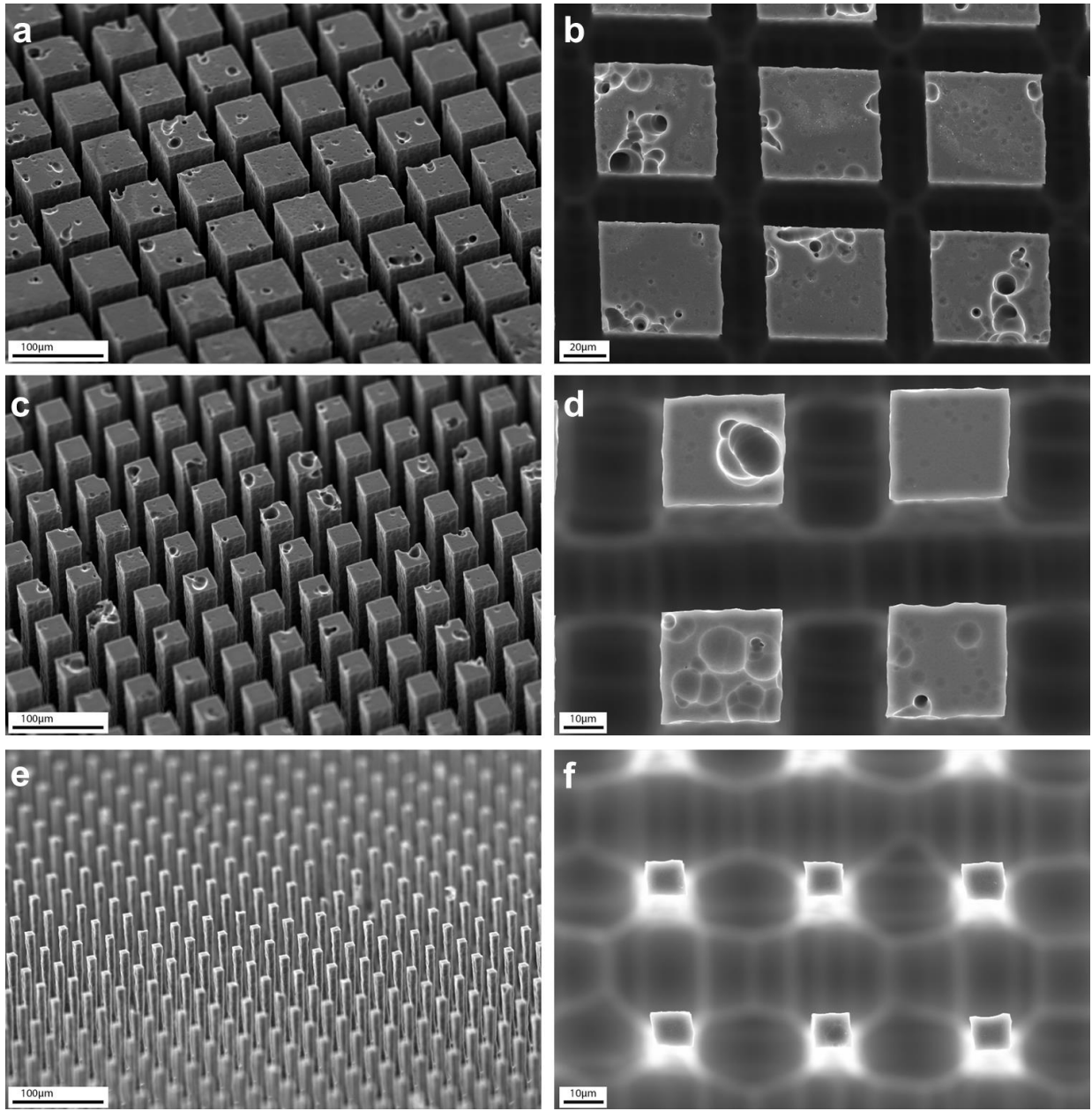

**Figure 2. SEM images of micro-pillar arrays (a, c, e) 40° tilt; Micro-pillar (b, d, f) top surface sizes: (a, b) 57 µm × 57 µm, (c, d) 31 µm × 31 µm, (e, f) 9 µm × 9 µm, and pulse energy 3 µJ.**

Figure 3 shows that the 32 µm x 32 µm micro-pillar array resulted in lowest reflectance among micro-pillar arrays, all of which exhibited significant light trapping activity as compared to the unstructured surface.

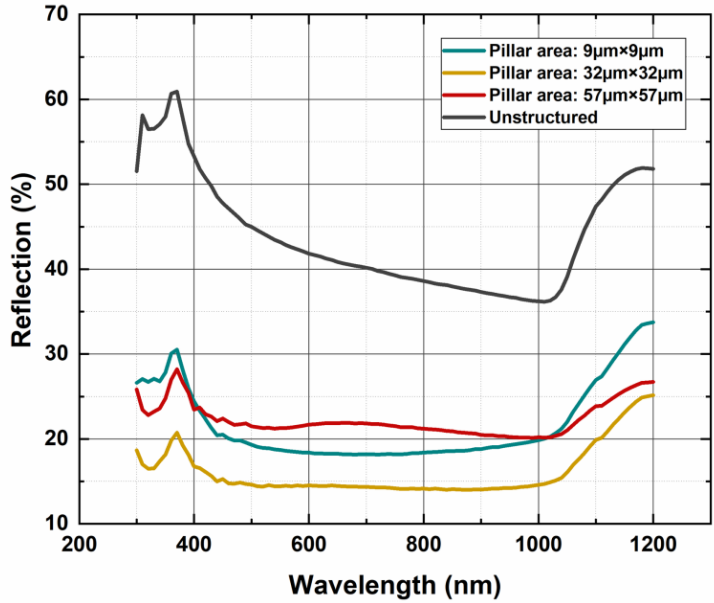

**Figure 3. Total reflectance of the fabricated Si micro-pillars with different areas and depths.**

## 3. Methods

In this study, the samples were prepared from double side polished, single-crystalline p-type Si wafers with <100> orientation. Two different sets of samples were prepared for the laser processing: (i) For subsurface modifications, the Si wafer had a thickness of 525 µm and a resistivity of 1-5 Ω.cm. (ii) For surface modifications, Si wafer with a thickness of 1000 µm and resistivity of 1-10 Ω.cm were used. With these two methods of preparation, we aimed to demonstrate an etching profile for both Cross-sectional Accessible Laser Processed (CALP) and Surface Accessible Laser Processed (SALP), which are discussed below. After the laser processing, the samples were treated with various etch recipes while systematically varying the components adopted from MEMC to determine the optimal composition.

### 3.1. Laser-Induced Si Modification

All Si modifications were done by 3D subsurface laser processing however, the extend of subsurface modifications toward the surface was different for different samples. The laser system was an in-house built nanosecond fiber laser, producing up to 5 W of average power at a wavelength of 1550 nm. The repetition rate of the laser was adjustable between 100 kHz to 1 MHz, limited by the generation of amplified spontaneous emission (ASE) at low repetition rate. In the experiments, the repetition rate of the laser was 100 kHz having a maximum pulse energy of 50 µJ and a pulse duration of ~5 ns. While we observed the desired modification at pulse energies higher than 2.5 µJ, we continued with 2.5 µJ pulse energies in our experiments. The laser beam was focused inside the samples by an aspheric lens (NA = 0.3, $f$ = 8mm). The polarization direction was along the vertical direction with respect to optical table during these experiments when the laser beam was parallel to optical table. We note that we did not observe any effect of polarization direction with respect to Si crystal orientation on the properties of the processed region. The samples with dimensions of 10 mm by 10 mm were mounted on a 3D motorized translation stage for scanning in all directions with sub-micron resolution.

Two different sets of samples were prepared for different opening access to the laser-modified region. For the first type, the laser was focused into the sample from the cross-sectional direction, and the sample was raster scanned laterally (in the *x-y* direction), as shown in Figure 4 (a). This scanning pattern forms a laser-modified plane parallel to the top surface inside the sample. The laser-modified thickness of each plane is ~4 μm, as confirmed by the IR camera images (Supporting video 1 – more details in ref. [28]). Several planes were created on top of each other, forming a laser-processed zone with a thickness of ~30 μm, as shown in Figure 4 (b). For this set of samples, the laser-modified zone is accessible to the etchant solutions from the cross-sectional surface at three sides with a 30-μm height openings with wafer dicing (Supporting Information 1). We will call this type of samples as CALP.

For the second type of the samples, the laser beam was focused inside the sample from the top surface and raster scanned in the *x-y* direction, as shown in Figure 4 (c). The cross-sectional thickness of one plane in the *z-direction* is ~80 μm. After the creation of each plane, the samples were shifted 20 μm down in the *z*-direction (corresponding to ~60 μm beam shift- up of the focal point inside of the samples due to the refractive index of Si), and the next modified plane was formed with ~20 μm overlap with the previous one. We repeated these steps until the last laser-modified plane was created on the top surface of the samples for a total depth of ~450 μm, as shown in Figure 4 (d). For this set of samples, the laser-modified zone is accessible to the etchant from the top surface (10 mm × 10 mm area). We will refer to these as SALP samples.

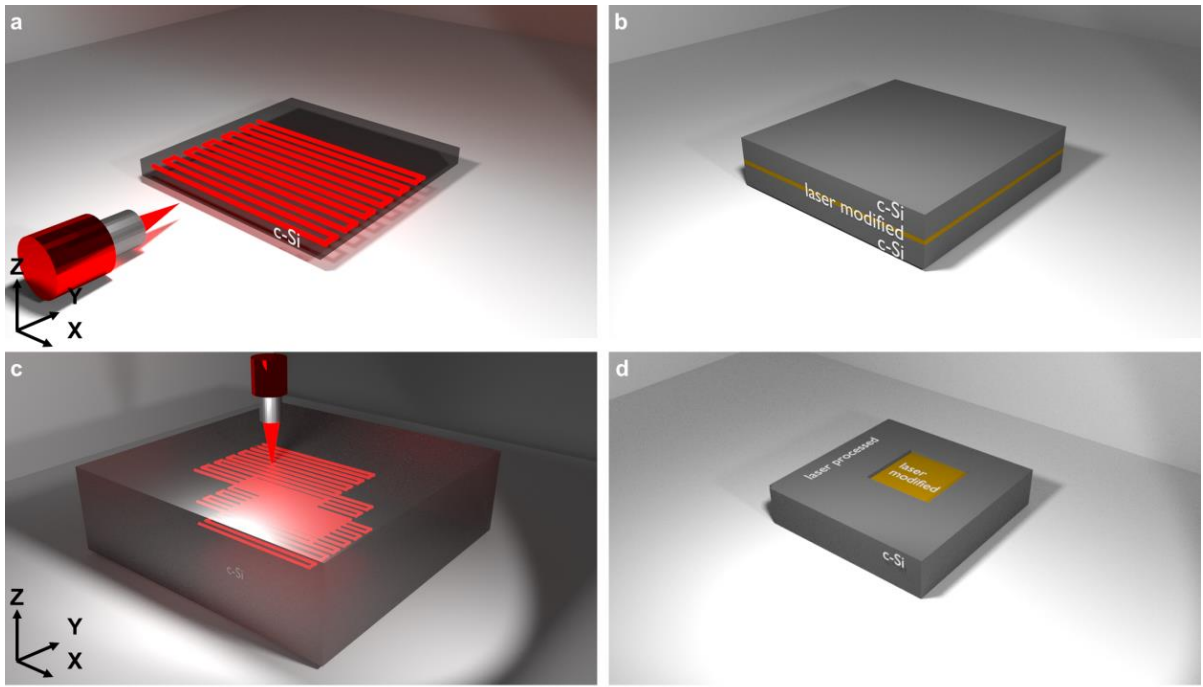

**Figure 4. (a, c) – Depiction of the laser-beam scanning direction (a) inside of c-Si for CALP samples, and (c) on SALP samples; (b, d) – 3D schematic representation of the laser processed zones for (b) CALP, and (d) for SALP.**

## 3.2. Chemical Etching

The effects of different etch solution mixtures are investigated on both CALP and SALP samples. Our motivation is studying selective etching of Si in CALP configuration since it represents a universal scenario in the 3D laser structuring of Si. This process does not necessarily create large laser processed surface areas where the etchant can quickly attack. It is essential to check the feasibility of the fabrication of 3D structures embedded inside the Si while heat transfer and diffusion constraints associated with narrow surface openings are present. As a complementary configuration, we introduced SALP structuring to create features from the surface up to the desired depth inside Si.

For the CALP structuring, the samples were diced into smaller individual samples with an area of 7.30 mm × 2.34 mm after laser modification. The dicing exposes three surfaces of the laser-modified region (one with 2.34 mm × 0.03 mm and two with 6.55 mm × 0.03 mm) to the etchant solution (see Figure 5 (a) and Figure S8). For the SALP structuring, the samples were diced with dimensions of 10 mm × 10 mm. The laser-modified region (2 mm × 2 mm) is exposed to the etchant from one side, vertically from the top surface (see Figure 5 (b)). Also, a set of unprocessed Si wafers were diced with identical sizes (reference samples). Before chemical treatments, all samples were subjected to standard cleaning 1 and 2 (SC 1, SC 2), which are used to remove organic contaminants as well as metallic and ionic contaminants from the surface of the samples, respectively [63].

As mentioned in the Introduction, MEMC solution is already used in the electronics industry for defect density determination of Si wafers. However, the MEMC solution offers a high etch rate at the expense of high selectivity and smooth surface morphology. In order to lend MEMC the preferred qualities in selectivity and surface morphology, we made a systematic analysis of the effect of each redox component [64], [65] on the etching chemistry. In order to investigate the effects of $HNO_3$ as an oxidant, copper (II) nitrate hydrate salt ($Cu(NO_3)_2.3H_2O$) as a catalyst [56], [66], and HF as an oxide dissolvent on selectivity, etch rate, and surface morphology, the concentration of each component was varied systematically. More details regarding the redox chemical reactions and catalytic activities can be found in Supporting Information 2. In this study, 12 different sets of solutions were considered. The concentrations of the components are separated into three subgroups which are highlighted in Table 1. Within each subset, we varied only one component of the mixture. The deionized wafer (DI $H_2O$) is also varied to keep the solution volume fixed to 100 ml.

**Table 1. Chemical composition of the solutions with different $Cu(NO_3)_2$, HF, and $HNO_3$ concentrations. The amount of DI $H_2O$ is adjusted to maintain 100 ml of solution corresponding to 100%.**

| | $Cu(NO_3)_2$ (g) | HF (vol%) | $HNO_3$ (vol%) | $CH_3COOH$ (vol%) |
|---|---|---|---|---|
| $Cu(NO_3)_2$ Comparison | 0.75 1.25 1.00 2.00 | 14.00 | 16.25 | 18.00 |
| HF Comparison | 1.00 | 7.00<br>10.00<br>12.00<br>14.00<br>16.00<br>18.00 | 16.25 | 18.00 |
| HNO$_3$ Comparison | 1.00 | 14.00 | 9.75<br>13.00<br>16.25<br>19.50<br>22.75 | 18.00 |

In addition to analyzing the effect of above-mentioned components of MEMC solution, we analyzed several alternative etching solutions to compare them with our etchant properties, including etch rate and selectivity. These results are presented in Supporting Information 3.

### 3.3. Etch Rate and Selectivity Redefined

The quantitative assessment of surface etching in 2D follows the established concepts of the etch rate and selectivity, as developed for semiconductor fabrication [67], [68]. When using lasers for etching, it seems that both etch rate (percentage of etched mass from modified region over time, i.e., the slope in the Figure 7 (a), (d), (g)) and selectivity (the ratio of etch rates of the laser-modified and unprocessed regions) need to be redefined since the process occurs in the subsurface, where etch thickness over time is not applicable. Therefore, we have redefined these parameters by using a 3D approach that necessitates using correction factors for the geometry.

In order to calculate the mass reduction due to the etching of the laser-modified regions in each chemical mixture (which is referred to as a Set), a pair of laser-processed and unprocessed c-Si samples were placed in the same solution as depicted in Figure 5. The mass reduction of the laser processed sample consists of two parts, that of the laser-modified region and the remaining (unprocessed) region of the same sample. We estimated the mass reduction of the unprocessed region of the laser-modified sample by measuring the mass reduction of the unprocessed sample. Then, by incorporating proper correction factors, we modeled and calculated the mass reduction of the laser-modified region (detailed in the Results and Discussion section). Various practically unmeasurable parameters are involved in these calculations in the static etching procedure. Those parameters include diffusion restriction for the mass and heat transports. Hence, proper estimation of the corresponding correction factors is essential (see Correction Factors Calculation & Etch Pit Radius Distribution Analysis). The SEM images taken from samples at the end of each etching process step presented in Supporting Information 4 are used to track and determine the smoothness of the surface morphology by analyzing etch pit size distribution.

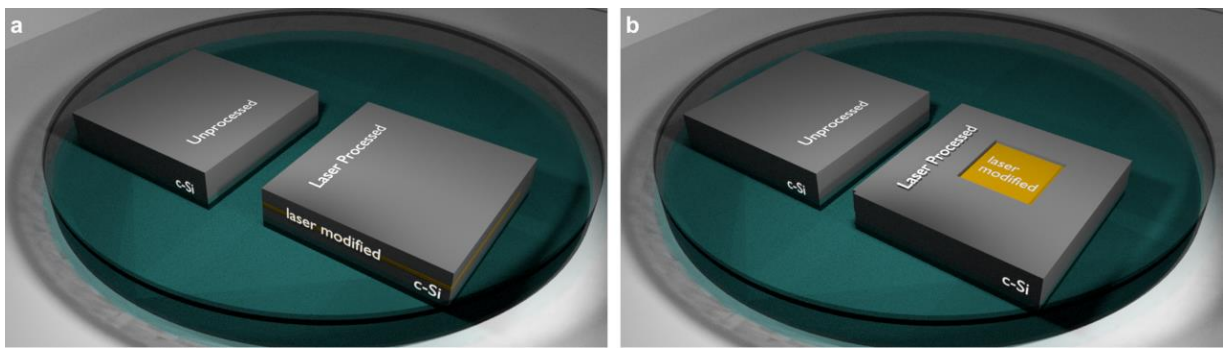

**Figure 5. Schematic representations of (a) CALP and (b) SALP samples in the chemical bath. In both cases, an unprocessed Si was added.**

In order to calculate the etch rates, mass reductions of the laser processed and the unprocessed samples were measured by weighting the samples after etching in each solution for 60 minutes with 10-minute intervals for CALP and 5-minute intervals for SALP. We specifically probed the interval between 10 to 60 minutes of etching durations since industrial applications favor short etch durations while too short durations may not be sufficient to fully etch modified regions. Based on the obtained etch rates, we calculate the selectivity of each solution set. Hence, we assess the effect of each component to determine the corresponding properties of the solution.

### 3.4. Density Functional Theory Calculations

Density functional theory (DFT) calculations were performed to understand the atomistic origins of the interaction between the etchants and the surfaces. All calculations presented here were conducted using the periodic, plane-wave-based Quantum Espresso code suite [69]. The exchange-correlation interaction was treated within the Perdew-Wang (PW-91) approximation [70], and electron-nucleus interaction was modeled using ultra soft pseudopotentials [71]. The plane-wave kinetic energy used to truncate the basis expansion was 40 Rydberg, and a vacuum distance of 14 Å was left between the images to reduce inter-image interaction. Atomic partial charges were calculated using the Bader charge decomposition as implemented by the Henkelman group [72].

According to the transmission electron microscopy (TEM) results from our previous study on a similar system [1] (presented in Supporting Information 5), the laser-modified regions appear to have a short-range order with a characteristic dimension of a few nanometers, interrupted by amorphous Si domains. The exact crystal structure of the ordered regions is not known; however, possible candidates are Si-I (FCC), Si-III (BCC) and Si-XII (Rhombohedral) phases, the latter two being known to form as a result of pressure or laser induced damage [73]. However, our Raman results do not show any indication of Si-III or Si-XII but rather a diminished Si-I peak. It is fair then to assume that the difference in reactivity behavior of the laser-processed and unprocessed regions stem from the amorphous Si domains. Furthermore, although dissolution mechanisms of amorphous Si and silicon dioxide ($SiO_2$) have been shown to be rather complicated and dependent on a variety of parameters [74], overall amorphous Si surfaces are expected to show larger reactivity with respect to those that are crystalline, due to undercoordination, and abundance of defect sites [75]. In our DFT calculations, we therefore model the surfaces of the processed regions by amorphous $SiO_2$ while the unmodified Si is modeled by oxidized and unoxidized reconstructed Si <100> surfaces. In the damaged portions of the samples, oxidation is expected to take place very rapidly. Therefore, we do not consider the unoxidized amorphous Si.

The pristine Si<100> surface was modeled by a slab of five layers with the two bottommost layers fixed to mimic bulk behavior. The Si <100> surface displayed several reconstructions to lower its surface energy, and a common reconstruction of this surface is the $2 \times 1$ or the dimer reconstruction [76], which was also applied in our calculations (Figure 11 (a)). Next, an oxygen-decorated Si surface was built by placing rows of O atoms on the surface (Figure 11 (b)). Finally, the amorphous $SiO_2$ simulation cell was constructed by extracting a 12 Å × 12 Å × 12 Å cube out of a previously conducted molecular dynamics (MD) investigation of a 3000-atom amorphous $SiO_2$ system [77]. In order to preserve stoichiometry, care was taken to include 36 Si and 72 O atoms. Once optimized in the bulk, a surface was formed via the extraction of the usual periodic slab geometry from bulk, and a DFT-based geometry optimization was performed on this surface slab. The density was found to be 2.1 g/cm$^3$, close to the experimental value of 2.2 g/cm$^3$.

## 4. Results and Discussion

To determine the effectiveness of each etching set, we have calculated the etch rate, etch pit radius distribution and selectivity of laser-modified structures for both CALP and SALP. In addition, the derivative of normalized etched mass and ratio of area normalized etched mass of the surface over time are calculated for SALP. The etch rate is defined as the mass reduction of the laser-modified region versus etching duration. The selectivity is defined as the etch rate of the laser-modified region divided by the etch rate of the unmodified region, indicating the rate at which the laser - modified region is etched away compared to the unmodified region. Note that it is also possible to use the sample volume instead of the mass since they only differ in a density factor which is almost same for both the laser-modified region and unprocessed samples. The etch pit radius distribution analysis, which is an indication of the surface porosity and roughness, was performed for different concentrations of the solution components. The SEM images (Supporting Information 4) show that the etched surfaces of laser-modified samples contain distinctive features such as etch pits in all considered sets. The observed etch pits on the surface are formed due to the interaction of the different solutions with the c-Si segments. Our etchant solution is formulated to reveal the defects and thus the laser-modified regions. The etchant attacks and enlarges the defects, including dopant impurities on the surface of the Si rather than attacking the perfect (unprocessed) Si lattice. By utilizing chemical etching and etch pit analysis, it is possible to calculate the defect density (counting the etch pits and dividing them by the surface area) that is grown in the crystal lattice. However, it is a destructive characterization technique for determining the crystal quality of crystalline samples.

### 4.1. Correction Factors Calculation & Etch Pit Radius Distribution Analysis

To calculate the etch rate and selectivity in terms of reduced mass of the samples, the mass of laser processed Si and the etched part must be determined. Thus, another unprocessed sample was used to find the mass change of the unmodified region. The mass difference between the laser-processed and the unprocessed samples is related to the laser-modified region. After each chemical process, the mass change of the unprocessed and processed samples was measured. The difference between the measured masses gives the etched mass of the laser-modified region, which needs to be corrected due to the complicated behavior of the etching process, coming from diffusion peculiarities of the mass and heat transports. Therefore, the mismatch of the initial masses before the chemical etching of unprocessed and laser processed samples, the volume etched inside the laser-modified region, and difference in the opening areas of crystalline region and laser-modified region of the processed sample being subjected to etchant need to be modeled accordingly. To calculate laser-modified mass correctly in CALP structures, we hypothesis three correction factors, namely, Correction Factor of Initial Mass (CFIM), Correction Factor of Etch Opening (CFEO), and Correction Factor of Attacked Area (CFAA).

To compare the SALP and CALP etching behaviors, the calculated mass reduction of the laser processed region is normalized to the opening areas exposed to the etching solution (0.46 mm$^2$ and 4 mm$^2$ for CALP and SALP, respectively). The detailed information regarding the correction factors and calculations for both SALP and CALP structures is given in Supporting Information 6. The measured weight losses are utilized in the analyses and the etched volumes obtained from etched openings (SEM images in Supporting Information 4) and etched area cross-sections (IR camera image in Supporting Information 7).

The etch pit size distribution is determined using the built-in image processing toolbox of MATLAB [78] and the SEM images of subsurface processed Si. It is worth mentioning that the subsurface and surface processing have an identical surface morphology. The histograms of etch pit size distribution for each material are presented in Effects of Solution Components section. The SEM image of the champion etchant is shown in Figure 6.

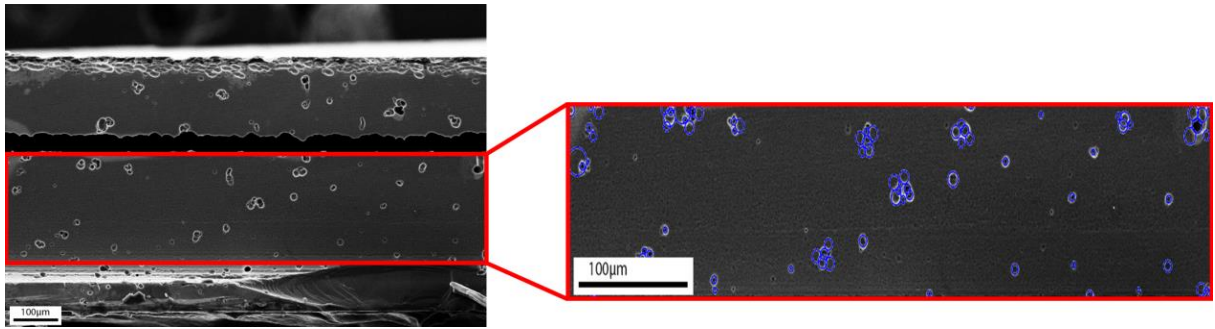

**Figure 6. The selected area for the champion etchant from cross-sectional SEM image and the result etch pit size distribution analysis.**

## 4.2. Effects of Solution Components on Etchant Chemistry

Etch rate and selectivity of the CALP structure, normalized etched mass and ratio of area normalized etched mass of the SALP structure, and etch pit radius distribution analysis for all chemical compositions were calculated based on the equations mentioned in Supporting Information 6. This information is categorized into three subgroups for each solution component. The etch rate for CALP and normalized etched mass for SALP structure is not constant across time since there is some latency before the etching starts. Thus, some delay may be required for the activation of Si-etchant reactions. Afterward, chemicals start to attack all regions until either the reactions saturate or the time frame ends. The etch rate shown in Figure 7 (a), (d), and (g) and normalized etched mass over time shown in Figure 7 (c), (f), and (i), and the selectivity of CALP shown in Figure 7 (b), (e), and (h) are presented for variation of Cu(NO$_3$)$_2$ concentrations from 0.75 to 2 g, HF concentrations from 7 ml to 18 ml, and HNO$_3$ concentrations from 9.75 to 22.75 ml in 100 ml of solution. Further analysis on the effects of solution components on SALP and CALP are presented in Supporting Information 8.

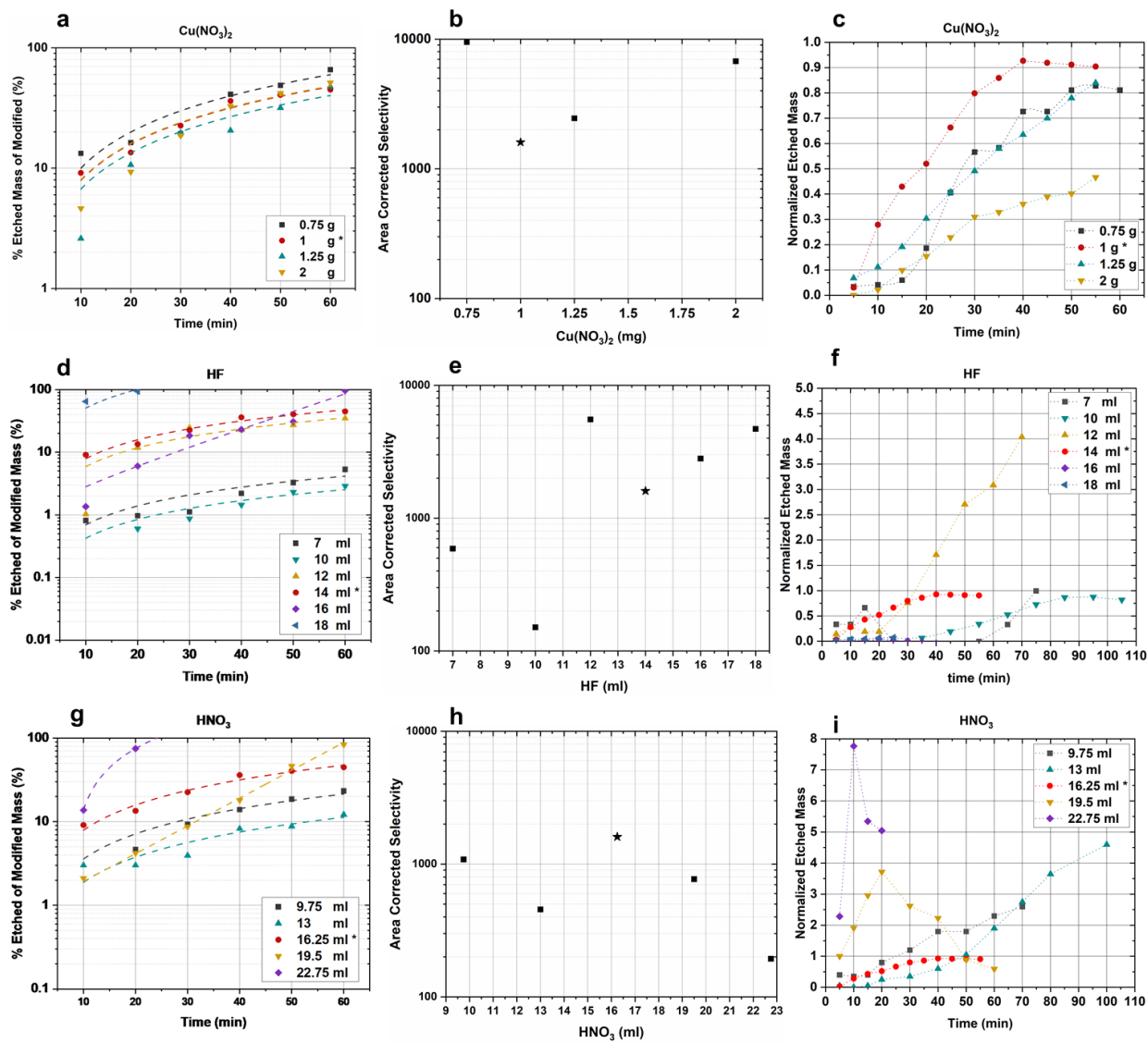

**Figure 7. (a), (d), (g)** Etch rate of laser-modified region over time for CALP. **(b), (e), (h)** Selectivity of laser-modified region compared to unprocessed sample for CALP. **(c), (f), (i)** Normalized etched mass of laser-modified region for SALP. The changing parameters are $Cu(NO_3)_2$ concentrations from 0.75 to 2 g (a), (b), (c); HF concentrations from 7 ml to 18 ml (d), (e), (f); $HNO_3$ concentrations from 9.75 to 22.75 ml in 100 ml of solution (g), (h), (i). The fixed parameters for each component variation are presented in Table 1. Our champion etchant concentration is marked by asterisk (*).

### 4.2.1. Effects of Copper Nitrate – Cu(NO3)2 Variation

The energy level of the $Cu^{2+}/Cu$ system is higher than the valence band (VB) of Si. For a given combination of $HF/HNO_3$, adding various amounts of $Cu(NO_3)_2$ corresponds to the Si surface being covered by copper nanoparticles (CuNPs) of different sizes. These metal particles also catalyzed the etching of Si in HF/HNO$_3$ solution, and their presence resulted in the selective attacks of modified regions. The Cu$^{2+}$ ions are reduced to solid Cu on the Si substrate and the oxidized surface is dissolved in HF. In addition, HNO$_3$ oxidizes the Si surface forming SiO$_2$ which is etched by the presence of HF.

In Figure 7 (a), it can be seen that the etch rate of the laser-modified region decreases as the amount of Cu(NO$_3$)$_2$ increases from 0.75 to 1.25 g. However, the etch rate starts to increase for a lower amount of Cu(NO$_3$)$_2$. In addition, as shown in Figure 7 (b), the selectivity for CALP shows a different trend where it is highest for 0.75 g of Cu(NO$_3$)$_2$ and drops drastically then increases by increasing the Cu(NO$_3$)$_2$ amount. In Figure 7 (c), the etched mass of the laser-modified region divided by total etched mass is highest where the Cu(NO$_3$)$_2$ amount is 1 g. However, the etched mass decreases as the amount of Cu(NO$_3$)$_2$ further increases from 0.75 to 2 g.

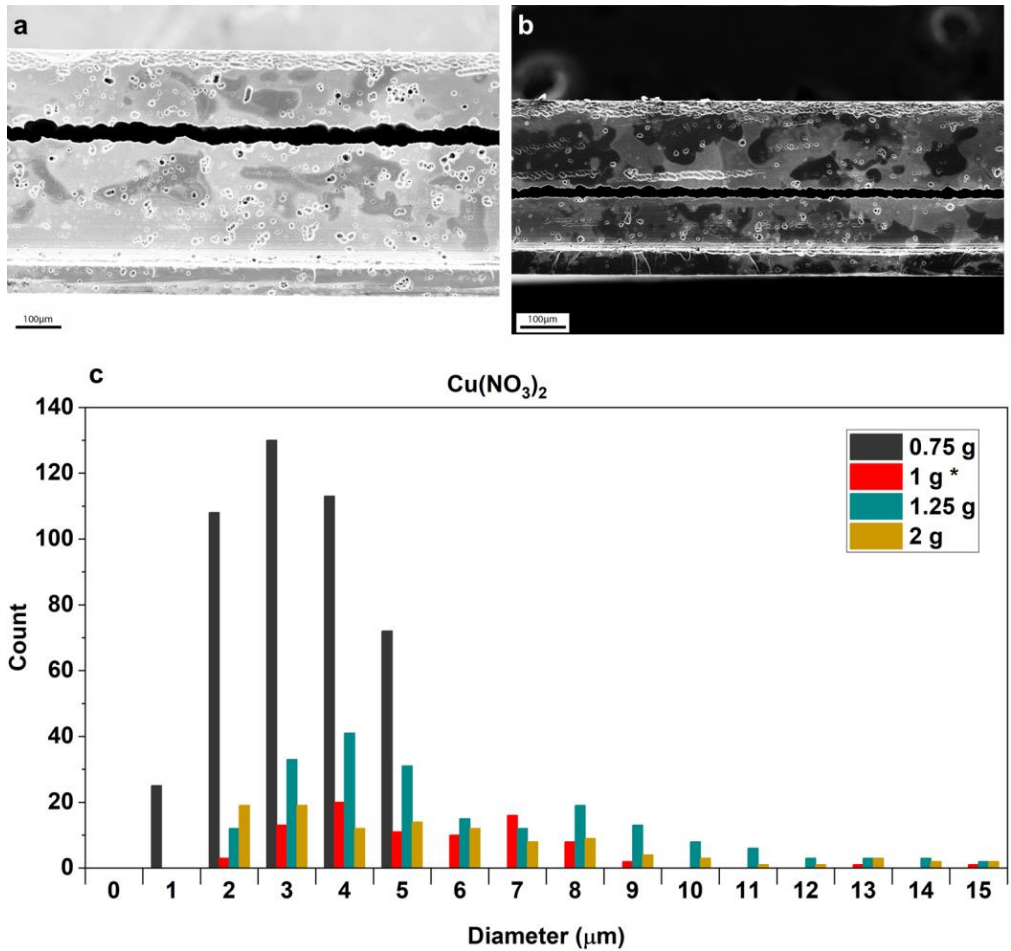

**Figure 8. SEM images of surface of samples having Cu(NO$_3$)$_2$ concentrations of (a) 0.75 g (b) 2 g in 100 ml of solution. (c) Etch pit radius distribution was obtained from SEM images by analyzing the surface morphology.**

The etch pit radius distribution, ranging between 1 – 15 µm, is identical for both SALP and CALP surfaces as shown in Figure 8 (c) which infer that the etching solution results in similar surface morphology regardless of the place of the laser modifications. The etch pit size distribution is in the order of $10^0$ µm. However, as the amount of $Cu(NO_3)_2$ increases, the resultant defects are found to have larger diameters. This can be attributed to the increase in the oxidation rate of the Si surface which increase the probability of revealing the defects embedded in the c-Si.

### 4.2.2.  Effects of Hydrofluoric Acid – HF Variation

Increasing the amount of HF beyond a certain point for a given ratio of Cu and $HNO_3$ hinders the suspension of the CuNPs. Thus, CuNPs will be adsorbed on the surface at random locations and promote the etching process in their vicinity. This process is fast and creates more holes, leading to a spongy Si surface morphology.

As can be seen in Figure 7 (d), a higher concentration of HF in the solution induces a very high etch rate and damages the sample, leading to poor surface morphology (Figure S4 (e), (f)). The effect of higher HF is more pronounced in laser-modified regions than the unprocessed one, resulting in higher selectivity. In Figure 7 (f), the etched mass of laser-modified divided by total etched mass region is well behaved for 14 ml of HF, showing a reasonable etch rate. However, the modified region is harshly attacked in the etching solution with high HF concentration due to the nonlinear diffusion of the defects.

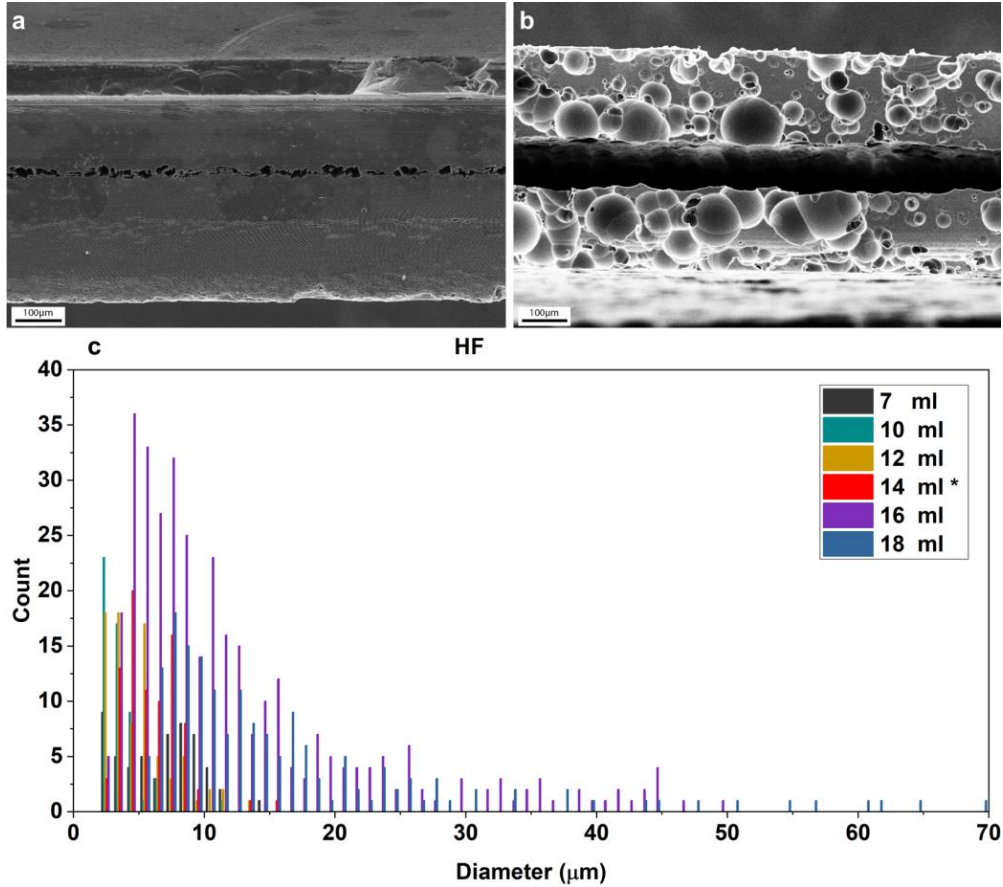

**Figure 9. SEM images of surface of samples having HF concentrations of (a) 7 ml (b) 18 ml in 100 ml of solution. (c) Etch pit radius distribution was obtained from SEM images by analyzing the surface morphology.**

In Figure 9 (c), the etch pit size distribution is below 15 µm when the HF concentration is moderate (7 ml to 14 ml). With a higher concentration of HF (above 16 ml), the etch pits exhibit larger defect sizes on the Si surfaces, as can be seen in the SEM image (Figure S4 (e), (f)).

### 4.2.3.   Effect of the oxidizing agent (Nitric Acid – HNO3) concentration

Cu ions are adsorbed on the Si surface. These ions attract minority negative charge carriers available in the p-type Si substrate. Thus, CuNPs collect electrons through Cu/Si interface and transfer them to the solution. The conglomeration speed of CuNPs depends on the rate of metal ion removal. So, if the formation of the CuNPs is slower than its removal, there will be high hole concentration on the surface that grows larger and larger. The amounts of oxidizing and reducing agents control how fast and aggressive Cu removes Si from the surface for a certain Cu amount. As a result, by fixing the amount of $Cu(NO_3)_2$ to 1 g, the combination of oxidation and reduction creates the intended structure. By tuning the oxidizing agent – $HNO_3$, it is possible to minimize the aggressiveness of the oxidation process in such a way that despite some distinct holes at the surface, the etchant does not wholly damage the Si surface and cause porosity. As discussed previously, increasing the $HNO_3$ controls the oxide formation on the CuNPs, which HF later etches. On the one hand, the presence of a high amount of $HNO_3$ in a given concentration of HF, i.e., 14 ml in 100 ml solution, causes CuNPs to become smaller. As a result, increasing $HNO_3$ makes the surface porous. On the other hand, the CuNPs must be maintained long enough to achieve the desired structure.

As shown in Figure 7 (g), (h), etch rates increase while selectivity decreases by increasing the $HNO_3$ concentration in the CALP structure. This confirms that increasing the amount of oxidizing agent in the solution increases the etch rate of the laser-modified region more than the unprocessed sample. This behavior is due to the generation of more excess hole-type carries at the surface, which accelerates the anodic chemical reactions. In Figure 7 (i), the etched mass of laser-modified sample divided by total etched mass is increased and harshly attacked on both regions by the increasing amount of $HNO_3$ concentration as the etching time increases.

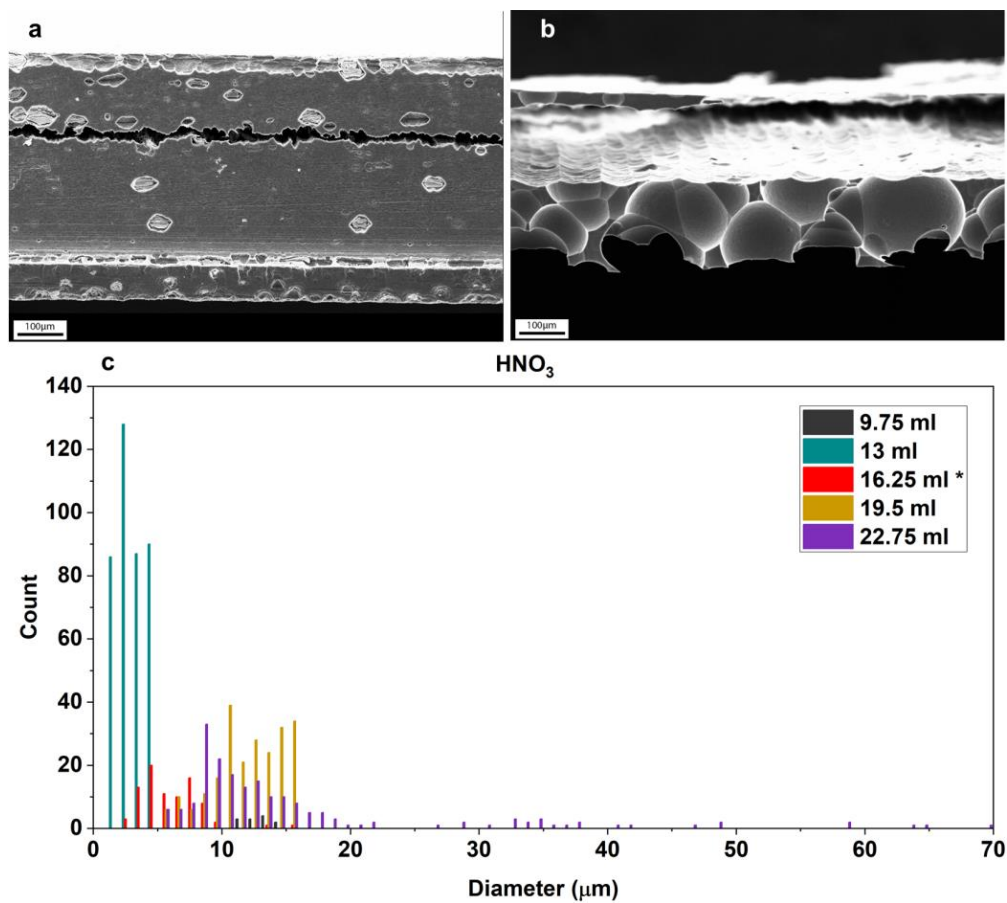

**Figure 10. SEM images of surface of samples having HNO₃ concentrations of (a) 9.75 ml (b) 22.75 ml in 100 ml of solution. (c) Etch pit radius distribution was obtained from SEM images by analyzing the surface morphology.**

The quality of the etched surface is a function of HNO₃, i.e., increasing the amount of HNO₃ will result in the formation of an engraved surface. Figure 10 (c) shows the etch pit radius size distribution is in the range of 1 – 70 µm except for the cases when HNO₃ concentration is high. The observed trends, particularly pertaining to the selectivity trends of the Cu and HF components, are corroborated with our DFT results presented in the next subsection.

The observed non-linear etching behavior in all etching solutions can be attributed to the different ratios of amorphous and polycrystalline Si phases in laser-modified region. These phases have different etching rates since amorphous is crystallographically disordered and polycrystalline is crystallographically locally ordered which is confirmed by TEM analysis (HRTEM, DAAF, SAED) presented in Supporting Information 5.

After surveying the abovementioned chemical compositions, we have successfully determined the optimal concentrations of HF:HNO₃:CH₃COOH:H₂O – 14.00:16.25:18.00:51.75 vol% with 0.01 g/ml Cu(NO₃)₂.3H₂O, which we refer to as *the Champion Etchant*. The enhanced composition yields more chemical treatment time required for specific applications. It is also dopant-type dependent and anisotropic (Supporting Information 9), which is beneficial for certain applications to etch laser-modified Si regions. The comparisons between MEMC and the enhanced composition in terms of orientation dependency and dopant type dependency are discussed in Supporting Information 9.

## 4.3. DFT Results

Interaction of the surface with HF was modeled by means of simultaneously placing a single H and a single F on the surface, whereas the interaction with Cu was represented by a single Cu atom on the surface (See Figure 11). Optimized geometries on the three surfaces can be seen in Figure 11. The adsorption energies were calculated using the following equation:

$$E_{\text{ads}} = E_{\text{tot}} - E_{\text{surf}} - E_{\text{mol}} \tag{1}$$

where $E_{\text{tot}}$ is the optimized energy of the entire system, $E_{\text{surf}}$ is the energy of the surface (Si<100>), oxygen-decorated Si<100>, and amorphous SiO$_2$) without the H/F or Cu atoms, and $E_{\text{mol}}$ is the energy of the adsorbates in isolation. In the case of the Cu adsorbate, $E_{\text{mol}}$ corresponds to the total energy of a single Cu atom in a large simulation cell. In contrast, in the case of H and F, it represents the sum of the energies of a single H and a single F atom, calculated in isolation separately. Among the several geometries explored, a few of the most stable optimized geometries are shown in Figure 11. The adsorption energies corresponding to the most stable geometries are shown in Table 2.

**Table 2. Adsorption energies of the H+F complex and the Cu atom.**

|  | **H+F** | **Cu** |
|---|---|---|
| Si<100> 2x1 reconstruction | -13.44 eV | -3.28 eV |
| Si<100> 2x1 reconstruction + O | -13.67 eV | -5.03 eV |
| Amorphous SiO$_2$ surface | -1.35 eV | -1.93 eV |

Several observations can be made based on these results. The amorphous SiO$_2$ surface appears to interact the weakest with the etchant species among the surfaces studied. While there is no difference in H/F adsorption energies between the Si and the oxygen-decorated Si surfaces, Cu appears to interact significantly more strongly with the Si-O surface. On the oxygen-decorated Si surfaces, when the H/F ions are adsorbed on the oxygen atoms, their adsorption energy is considerably lower than partial adsorption on the Si atoms. On the other hand, Cu appears to benefit from the interaction with the oxygen atoms. In order to verify the ionic nature of the H and F atoms, the partial charges were calculated and found to be about -1.0 eV and +1.0 eV for F and H adsorbates, respectively. The Bader charge of the Cu atom varies between 0 and +0.3 eV depending on the adsorption site. This is consistent with the experimental insight that Cu clusters are first positively charged before reduction.

The HF component displays no selectivity between oxidized and pristine crystalline Si, as evidenced by the identical adsorption energies of the H and F ions. The adsorption energy of the HF unit on the c-Si surface is one order of magnitude larger than the amorphous surface. This significant difference between the adsorption energies explains the selectivity of the HF etchant towards the laser-processed surface. Etching of the surface proceeds through the desorption of adsorbed species. Therefore, the lower adsorption energy in the case of the amorphous oxide likely points towards facile desorption as well.

This trend continues with the catalyst component, namely the Cu atom. Cu adsorption on the oxidized c-Si surface is much larger than the a-SiO$_2$ considered. Since the Cu component should ideally desorb before it has time to form agglomerates, the significantly lower adsorption energy on the amorphous SiO$_2$ surface appears to be much more advantageous than the crystal surfaces.

In particular, the Cu adsorption energy on the oxidized c-Si surface is particularly high, and the positive charge is larger. This behavior likely increases the dwelling time of Cu on the surface and encourages agglomeration, deactivating the catalyst.

We emphasize in passing one more time that the crystalline portions of the laser-processed domains appear to be a combination of nanocrystalline domains (Supporting Information 10) interrupted by grain boundaries with amorphous phase. Therefore, the difference in dissolution behavior of the processed and unprocessed regions is dictated by these portions.

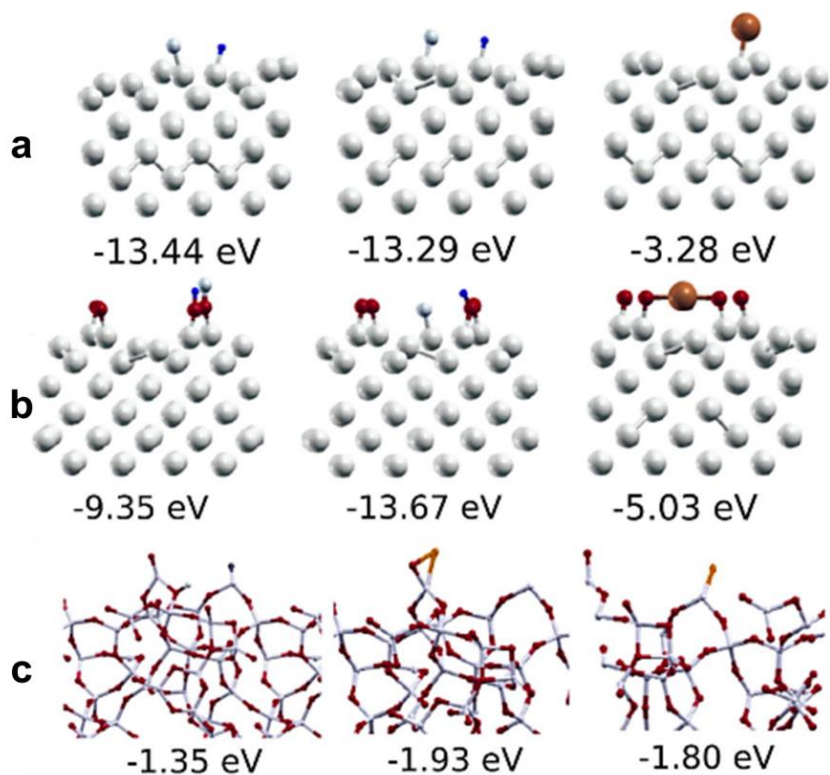

**Figure 11. Adsorption of HF and Cu on (a) reconstructed Si <100> surface; (b) oxygen-decorated Si <100> surface; (c) amorphous SiO2 surface. The Si, O, H, F, and Cu atoms are represented by white, red, gray, blue, and bronze spheres, respectively. The adsorption energies are given below each figure.**

## 5. Conclusion

The optimization of a $Cu(NO_3)_2$-based etching solution on the novel selective etching of surface and subsurface laser-modified regions (CALP and SALP) is presented. The laser modifications are done by utilizing a nanosecond IR fiber laser on p-type c-Si wafers. The laser-modified samples are treated with a number of MEMC-like etchants in which the chemical component concentrations are varied in order to survey for an especially optimal etchant for laser-modified Si. For investigating the effect of each component, different sets of solutions are prepared by changing the concentrations. To obtain the etch rates, the mass difference of laser-modified and unprocessed samples is measured up to 60 minutes. By applying correction factors of initial mass, etch opening, and attack area, the mass reduction of the laser-modified region and unprocessed c-Si samples are calculated. The selectivity of each set is calculated as the ratio of laser-modified region and unprocessed c-Si samples. The c-Si smoothness is represented by calculating etch pit radius size distribution which was obtained from SEM images after 60 minutes. The etch rate, selectivity, and surface morphology of subsurface laser-modified regions are well controlled during the experiments by varying the etching concentration of the etching solution components. The selection of the component concentrations depends on the desired applications since a different combination of etch rate, selectivity, and surface morphology can be achieved. For instance, decreasing $Cu(NO_3)_2$ to 0.75 g in 100 ml solution leads to fast and selective etching at the cost of c-Si surface smoothness. The etching rates show nonlinear behavior due to presence of both amorphous and polycrystalline phases in the laser-modified regions confirmed by TEM. The adsorption energy calculations performed within the DFT method clearly corroborate the experimental findings. Calculated HF and Cu adsorption energies display significant diversity on the crystalline and amorphous surfaces. Since adsorption energy is a good indicator for reactivity, the difference in the adsorption energies can be interpreted as a potential explanation for the selectivity trends. In addition, decreasing $HNO_3$ leads to a low etch rate and selectivity while providing smooth c-Si surface morphology. Some applications which may use small samples require high selectivity but a low etch rate which can be achieved by increasing the $Cu(NO_3)_2$ to 2 g in 100 ml solution. After an extensive survey of component concentrations, we achieved an optimum composition of $HF:HNO_3:CH_3COOH:H_2O - 14.00:16.25:18.00:51.75$ vol% with 0.01 g/ml $Cu(NO_3)_2.3H_2O$ as *the champion etchant* for surface and subsurface laser-modified Si which shows more than 1600 selectivity as well as etch pit sizes in the range of $1-10$ µm indicating relatively smooth and low defective surface. In addition, by comparing SALP and CALP results, it is concluded that surface etching has a higher etch rate which is expectable due to heat transfer and diffusion restrictions in the subsurface samples. In order to demonstrate the capability of this especially optimized etchant, a micro-pillar array is fabricated with a controlled size/depth combination in a two-step process. With this technique, micro-pillars with lateral dimensions as small as 9 µm × 9 µm and with a depth as large as 118 µm were successfully fabricated yielding an aspect ratio >13. We successfully utilized our novel technique to fabricate periodic arrays of pillars with a controlled aspect ratio as a light-trapping structure for use in photonic devices such as photon detectors and solar cells. Direct writing of 3D Si structures is expected to attract increasing attention in advanced device fabrication technologies including, but not limited to MEMS.

## Acknowledgments

M.Z.B. and B.R contributed equally to this work. This work was supported by the Scientific and Technological Research Council of Turkey (TÜBİTAK) under grant nr. 113M931 (I.P.), 118E995 (A.B.) and 20AG024 (R.T.); the European Research Council (ERC) through the NLL (grant nr. 617521), and the SUPERSONIC (grant nr. 966846) projects (F.Ö.İ.); the Turkish Academy of Sciences Young Investigator Program (TUBA-GEBIP) Award (O.T.). B.R. acknowledges the financial support of the Academy of Finland (grant nr. 331313). We kindly thank Dr. Mehmet Koç of ODTÜ-GÜNAM for help with some of the figures.

## References

[1]     O. Tokel *et al.*, "In-chip microstructures and photonic devices fabricated by nonlinear laser lithography deep inside silicon," *Nat. Photonics*, vol. 11, no. 10, pp. 639–645, 2017.

[2]     I. Pavlov, E. Dülgergil, E. Ilbey, and F. Ö. Ilday, "10 W, 10 ns, 50 kHz all-fiber laser at 1.55 µm," in *Conference on Lasers and Electro-Optics 2012*, 2012, p. CTu2M.5.

[3]     A. H. Nejadmalayeri, P. R. Herman, J. Burghoff, M. Will, S. Nolte, and A. Tünnermann, "Inscription of

optical waveguides in crystalline silicon by mid-infrared femtosecond laser pulses," *Opt. Lett.*, vol. 30, no. 9, p. 964, 2005.

[4]  B. Öktem *et al.*, "Nonlinear laser lithography for indefinitely large-area nanostructuring with femtosecond pulses," *Nat. Photonics*, vol. 7, no. 11, pp. 897–901, 2013.

[5]  M. Chambonneau, D. Grojo, O. Tokel, F. Ö. Ilday, S. Tzortzakis, and S. Nolte, "In-Volume Laser Direct Writing of Silicon—Challenges and Opportunities," *Laser Photon. Rev.*, vol. 15, no. 11, p. 2100140, Nov. 2021.

[6]  L. Jiang, A. D. Wang, B. Li, T. H. Cui, and Y. F. Lu, "Electrons dynamics control by shaping femtosecond laser pulses in micro/nanofabrication: Modeling, method, measurement and application," *Light Sci. Appl.*, vol. 7, no. 2, pp. 1–27, 2018.

[7]  H. Tang *et al.*, "Experimental quantum fast hitting on hexagonal graphs," *Nat. Photonics*, vol. 12, no. 12, pp. 754–758, 2018.

[8]  H. Tang *et al.*, "Experimental two-dimensional quantum walk on a photonic chip," *Sci. Adv.*, vol. 4, no. 5, pp. 1–6, 2018.

[9]  K. Kumar, K. K. Lee, J. Li, J. Nogami, N. P. Kherani, and P. R. Herman, "Quantized structuring of transparent films with femtosecond laser interference," *Light Sci. Appl.*, vol. 3, no. November 2013, pp. 1–7, 2014.

[10]  R. Osellame, H. J. W. M. Hoekstra, G. Cerullo, and M. Pollnau, "Femtosecond laser microstructuring: An enabling tool for optofluidic lab-on-chips," *Laser Photonics Rev.*, vol. 5, no. 3, pp. 442–463, 2011.

[11]  R. G. H. Van Uden *et al.*, "Ultra-high-density spatial division multiplexing with a few-mode multicore fibre," *Nat. Photonics*, vol. 8, no. 11, pp. 865–870, 2014.

[12]  C. Hnatovsky, R. S. Taylor, E. Simova, V. R. Bhardwaj, D. M. Rayner, and P. B. Corkum, "Polarization-selective etching in femtosecond laser-assisted microfluidic channel fabrication in fused silica," *Opt. Lett.*, vol. 30, no. 14, p. 1867, 2005.

[13]  K. Sugioka *et al.*, "Femtosecond laser 3D micromachining: A powerful tool for the fabrication of microfluidic, optofluidic, and electrofluidic devices based on glass," *Lab Chip*, vol. 14, no. 18, pp. 3447–3458, 2014.

[14]  M. S. Giridhar, K. Seong, A. Schülzgen, P. Khulbe, N. Peyghambarian, and M. Mansuripur, "Femtosecond pulsed laser micromachining of glass substrates with application to microfluidic devices," *Appl. Opt.*, vol. 43, no. 23, pp. 4584–4589, 2004.

[15]  R. T. and A. B. Mona Zolfaghari Borra, Hisham Nasser, Emine Hande Ciftpinar, Ahmet Turnali, Petro Deminskyi, Tahir Colakoglu, Onur Tokel, Fatih Omer Ilday, Ihor Pavlov, "Slicing Crystalline Silicon Wafer by Deep Subsurface Laser Processing and Selective Chemical Etching," 2019.

[16]  Mona, "IMPLEMENTATION OF STRONG LIGHT-MATTER INTERACTION FOR FABRICATION AND LIGHT MANAGEMENT OF THIN CRYSTAL SILICON SOLAR CELLS," Middlee East Technical University, 2021.

[17]  E. Granados, M. Martinez-Calderon, M. Gomez, A. Rodriguez, and S. M. Olaizola, "Photonic structures in diamond based on femtosecond UV laser induced periodic surface structuring (LIPSS)," *Opt. Express*, vol. 25, no. 13, p. 15330, 2017.

[18]  J. Bonse *et al.*, "Applications of laser-induced periodic surface structures (LIPSS)," *Laser-based Micro-Nanoprocessing XI*, vol. 10092, no. February 2017, p. 100920N, 2017.

[19]  S. Dottermusch *et al.*, "Micro-cone textures for improved light in-coupling and retroreflection-inspired light trapping at the front surface of solar modules," *Prog. Photovoltaics Res. Appl.*, vol. 27, no. 7, pp. 593–602, 2019.

[20]  B. Radfar, F. Es, and R. Turan, "Effects of different laser modified surface morphologies and post-texturing cleanings on c-Si solar cell performance," *Renew. Energy*, vol. 145, pp. 2707–2714, 2020.

[21] J. Ding *et al.*, "A laser texturing study on multi-crystalline silicon solar cells," *Sol. Energy Mater. Sol. Cells*, vol. 214, no. May, p. 110587, 2020.

[22] S. Kontermann, T. Gimpel, A. L. Baumann, K. M. Guenther, and W. Schade, "Laser processed black silicon for photovoltaic applications," *Energy Procedia*, vol. 27, pp. 390–395, 2012.

[23] M. Otto *et al.*, "Black silicon photovoltaics," *Adv. Opt. Mater.*, vol. 3, no. 2, pp. 147–164, 2015.

[24] X. Liu, B. Radfar, K. Chen, T. P. Pasanen, V. Vähänissi, and H. Savin, "Tailoring Femtosecond-Laser Processed Black Silicon for Reduced Carrier Recombination Combined with >95% Above-Bandgap Absorption," *Adv. Photonics Res.*, p. 2100234, Jan. 2022.

[25] V. Y. Fedorov, M. Chanal, D. Grojo, and S. Tzortzakis, "Accessing Extreme Spatiotemporal Localization of High-Power Laser Radiation through Transformation Optics and Scalar Wave Equations," *Phys. Rev. Lett.*, vol. 117, no. 4, pp. 1–5, 2016.

[26] M. Chambonneau, Q. Li, M. Chanal, N. Sanner, and D. Grojo, "Writing waveguides inside monolithic crystalline silicon with nanosecond laser pulses," *Opt. Lett.*, vol. 41, no. 21, p. 4875, Nov. 2016.

[27] M. Chanal, V. Y. Fedorov, M. Chambonneau, R. Clady, S. Tzortzakis, and D. Grojo, "Crossing the threshold of ultrafast laser writing in bulk silicon," *Nat. Commun.*, vol. 8, no. 1, pp. 1–6, 2017.

[28] G. C. Giakos *et al.*, "Laser imaging through scattering media," *Conf. Rec. - IEEE Instrum. Meas. Technol. Conf.*, vol. 1, pp. 433–437, 2004.

[29] K. M. Davis, K. Miura, N. Sugimoto, and K. Hirao, "Writing waveguides in glass with a femtosecond laser," *Opt. Lett.*, vol. 21, no. 21, p. 1729, 1996.

[30] S. Jesse, A. J. Pedraza, J. D. Fowlkes, and J. D. Budai, "Etching-enhanced ablation and the formation of a microstructure in silicon by laser irradiation in an SF6 atmosphere," *J. Mater. Res.*, vol. 17, no. 5, pp. 1002–1013, 2002.

[31] T. H. Her, R. J. Finlay, C. Wu, S. Deliwala, and E. Mazur, "Microstructuring of silicon with femtosecond laser pulses," *Appl. Phys. Lett.*, vol. 73, no. 12, pp. 1673–1675, 1998.

[32] D. Riedel, J. L. Hernandez-Pozos, R. E. Palmer, and K. W. Kolasinski, "Fabrication of ordered arrays of silicon cones by optical diffraction in ultrafast laser etching with SF6," *Appl. Phys. A Mater. Sci. Process.*, vol. 78, no. 3, pp. 381–385, 2004.

[33] D. Mills and K. W. Kolasinski, "Laser-etched silicon pillars and their porosification," *J. Vac. Sci. Technol. A Vacuum, Surfaces, Film.*, vol. 22, no. 4, pp. 1647–1651, 2004.

[34] D. Mills and K. W. Kolasinski, "Non-lithographic method of forming ordered arrays of silicon pillars and macropores," *J. Phys. D. Appl. Phys.*, vol. 38, no. 4, pp. 632–636, 2005.

[35] paper cm_2_6. R. A. Sabet, A. Ishraq, and O. Tokel, "Laser nanofabrication deep inside silicon wafers," in 2021 Conference on Lasers and Electro-Optics Europe and European Quantum Electronics Conference, OSA Technical Digest (Optica Publishing Group, 2021), "Laser nano-fabrication deep buried inside silicon wafers," 2021.

[36] O. T. R. A. Sabet, A. Saltık, "Creating efficient nano-gratings buried in silicon through laser nano-lithography," 2023.

[37] S. W. Bedell, D. K. Sadana, K. Fogel, H. Chen, and A. Domenicucci, "Quick turnaround technique for highlighting defects in thin Si/SiGe bilayers," *Electrochem. Solid-State Lett.*, vol. 7, no. 5, pp. 105–107, 2004.

[38] B. Miao *et al.*, "Improved metal assisted chemical etching method for uniform, vertical and deep silicon structure," *J. Micromechanics Microengineering*, vol. 27, no. 5, 2017.

[39] L. Romano *et al.*, "Metal assisted chemical etching of silicon in the gas phase: a nanofabrication platform for X-ray optics," *Nanoscale Horizons*, vol. 5, no. 5, pp. 869–879, 2020.

[40] N. Van Toan, M. Toda, T. Hokama, and T. Ono, "Cantilever with High Aspect Ratio Nanopillars on Its Top

Surface for Moisture Detection in Electronic Products," *Adv. Eng. Mater.*, vol. 19, no. 11, pp. 1–5, 2017.

[41] M. Z. Mohd Zin, E. H. Felix, Y. Wahab, and M. N. Bakar, "Process development and characterization towards microstructural realization using laser micromachining for MEMS," *SN Appl. Sci.*, vol. 2, no. 5, pp. 1–5, 2020.

[42] R. Dumke, M. Volk, T. Müther, F. B. J. Buchkremer, G. Birkl, and W. Ertmer, "Micro-optical Realization of Arrays of Selectively Addressable Dipole Traps: A Scalable Configuration for Quantum Computation with Atomic Qubits," *Phys. Rev. Lett.*, vol. 89, no. 9, pp. 3–6, 2002.

[43] R. Soref, "Mid-infrared photonics in silicon and germanium," *Nat. Photonics*, vol. 4, no. 8, pp. 495–497, 2010.

[44] Z. Fang, Q. Y. Chen, and C. Z. Zhao, "A review of recent progress in lasers on silicon," *Opt. Laser Technol.*, vol. 46, no. 1, pp. 103–110, 2013.

[45] X. Li *et al.*, "Improvement of saw damage removal to fabricate uniform black silicon nanostructure on large-area multi-crystalline silicon wafers," *Sol. Energy*, vol. 204, no. May, pp. 577–584, 2020.

[46] X. Wei, Z. Xiao, Z. Yue, H. Huang, and L. Zhou, "Texturization of diamond wire sawn multi-crystalline silicon wafers by micro-droplet etching," *Mater. Sci. Semicond. Process.*, vol. 115, no. February, p. 105075, 2020.

[47] M. Zolfaghari Borra *et al.*, "A feasibility study for controlling self-organized production of plasmonic enhancement interfaces for solar cells," in *Applied Surface Science*, 2014, vol. 318, pp. 43–50.

[48] T. E. Scheul, E. Khorani, T. Rahman, M. D. B. Charlton, and S. A. Boden, "Wavelength and angle resolved reflectance measurements of pyramidal textures for crystalline silicon photovoltaics," *Prog. Photovoltaics Res. Appl.*, no. January, pp. 1–10, 2020.

[49] H. Nasser, M. Z. Borra, E. H. Çiftpınar, B. Eldeeb, and R. Turan, "Fourteen percent efficiency ultrathin silicon solar cells with improved infrared light management enabled by hole-selective transition metal oxide full-area rear passivating contacts," *Prog. Photovoltaics Res. Appl.*, Nov. 2021.

[50] F. Es, G. Baytemir, M. Kulakci, and R. Turan, "Metal-assisted nano-textured solar cells with SiO2/Si3N4 passivation," *Sol. Energy Mater. Sol. Cells*, vol. 160, no. November 2016, pp. 269–274, 2017.

[51] T. H. Fung *et al.*, "Improved emitter performance of RIE black silicon through the application of in-situ oxidation during POCl3 diffusion," *Sol. Energy Mater. Sol. Cells*, vol. 210, no. February, 2020.

[52] B.Schwartz and H.Robbins, "Chemicaletchofsilicon.Pdf." J.Electrochem.SOC, 1976.

[53] W. C. Dash, "Copper precipitation on dislocations in silicon," *J. Appl. Phys.*, vol. 27, no. 10, pp. 1193–1195, 1956.

[54] A. Sirtl, E., & Adler, "Chromic acid-hydrofluoric acid as specific reagents for the development of etching pits in silicon," *Zeitschrift für Met.*, vol. 52, pp. 529–534, 1961.

[55] H. Seiter, "INTEGRATIONAL ETCHING METHODS," in *Semiconductor Silicon/1977*, 1977, p. 187.

[56] T. C. Chandler, "MEMC Etch—A Chromium Trioxide-Free Etchant for Delineating Dislocations and Slip in Silicon," *J. Electrochem. Soc.*, vol. 137, no. 3, pp. 944–948, 1990.

[57] Y. Saito and Y. Matsushita, "Etching solution for evaluating crystal faults," EP0281115B1, 1994.

[58] National Institute for Occupational Safety and Health (NIOSH), "NIOSH chemical carcinogen policy," *Curr. Intell. Bull. 68, DHHS Publ. Number 2017-100*, pp. 1–28, 2017.

[59] E. Garnett and P. Yang, "Light trapping in silicon nanowire solar cells," *Nano Lett.*, vol. 10, no. 3, pp. 1082–1087, 2010.

[60] Y. Wang, L. Yang, Y. Liu, Z. Mei, W. Chen, and J. Li, "Maskless inverted pyramid texturization of silicon," *Nat. Publ. Gr.*, no. 0316, pp. 1–6, 2015.

[61]     S. Greco, S. Dal Zilio, A. Bek, M. Lazzarino, and D. Naumenko, "Frequency Modulated Raman Spectroscopy," *ACS Photonics*, vol. 5, no. 2, pp. 312–317, Feb. 2018.

[62]     D. Naumenko, V. Toffoli, S. Greco, S. Dal Zilio, A. Bek, and M. Lazzarino, "A micromechanical switchable hot spot for SERS applications," *Appl. Phys. Lett.*, vol. 109, no. 13, p. 131108, Sep. 2016.

[63]     W. Kern, "The Evolution of Silicon Wafer Cleaning Technology," *J. Electrochem. Soc.*, vol. 137, no. 6, pp. 1887–1892, Jun. 1990.

[64]     K. Peng *et al.*, "Fabrication of single-crystalline silicon nanowires by scratching a silicon surface with catalytic metal particles," *Adv. Funct. Mater.*, vol. 16, no. 3, pp. 387–394, 2006.

[65]     Z. P. Huang, N. Geyer, L. F. Liu, M. Y. Li, and P. Zhong, "Metal-assisted electrochemical etching of silicon," *Nanotechnology*, vol. 21, no. 46, 2010.

[66]     Y. T. Lu and A. R. Barron, "Anti-reflection layers fabricated by a one-step copper-assisted chemical etching with inverted pyramidal structures intermediate between texturing and nanopore-type black silicon," *J. Mater. Chem. A*, vol. 2, no. 30, pp. 12043–12052, 2014.

[67]     X. Leng, C. Wang, and Z. Yuan, "Progress in metal-assisted chemical etching of silicon nanostructures," *Procedia CIRP*, vol. 89, pp. 26–32, 2020.

[68]     L. Yang *et al.*, "The Fabrication of Micro/Nano Structures by Laser Machining," *Nanomaterials*, vol. 9, no. 12, p. 1789, Dec. 2019.

[69]     P. Giannozzi *et al.*, "QUANTUM ESPRESSO: A modular and open-source software project for quantum simulations of materials," *J. Phys. Condens. Matter*, vol. 21, no. 39, 2009.

[70]     J. P. Perdew, "Unified theory of exchange and correlation beyond the local density approximation," *Electron. Struct. Solids '91*, vol. 17, pp. 11–20, 1991.

[71]     D. Vanderbilt, "Soft self-consistent pseudopotentials in a generalized eigenvalue formalism," *Phys. Rev. B*, vol. 41, no. 11, pp. 7892–7895, Apr. 1990.

[72]     W. Tang, E. Sanville, and G. Henkelman, "A grid-based Bader analysis algorithm without lattice bias," *J. Phys. Condens. Matter*, vol. 21, no. 8, 2009.

[73]     N. Fujisawa, J. S. Williams, and M. V. Swain, "On the cyclic indentation behavior of crystalline silicon with a sharp tip," *J. Mater. Res.*, vol. 22, no. 11, pp. 2992–2997, Nov. 2007.

[74]     X. Chen, C. Liu, J. Ke, J. Zhang, X. Shu, and J. Xu, "Subsurface damage and phase transformation in laser-assisted nanometric cutting of single crystal silicon," *Mater. Des.*, vol. 190, p. 108524, May 2020.

[75]     P. M. Dove, N. Han, A. F. Wallace, and J. J. De Yoreo, "Kinetics of amorphous silica dissolution and the paradox of the silica polymorphs," *Proc. Natl. Acad. Sci.*, vol. 105, no. 29, pp. 9903–9908, Jul. 2008.

[76]     C. Yang and H. Chuan Kang, "Geometry of dimer reconstruction on the C(100), Si(100), and Ge(100) surfaces," *J. Chem. Phys.*, vol. 110, no. 22, pp. 11029–11037, 1999.

[77]     S. Le Roux and V. Petkov, "ISAACS – interactive structure analysis of amorphous and crystalline systems," *J. Appl. Crystallogr.*, vol. 43, no. 1, pp. 181–185, Feb. 2010.

[78]     Natick, "MATLAB." Massachusetts: The MathWorks Inc, 2021.

# Supporting Information

## Supporting Information 1: Dicing of CALP Samples

In the CALP samples, the laser processing modifies the Si below the surface. The total area of processed region was 10 mm×10 mm. In order to increase the number of samples for more investigations, the samples were diced to the size of 7.30 mm×2.34 mm which is shown in Figure S1.

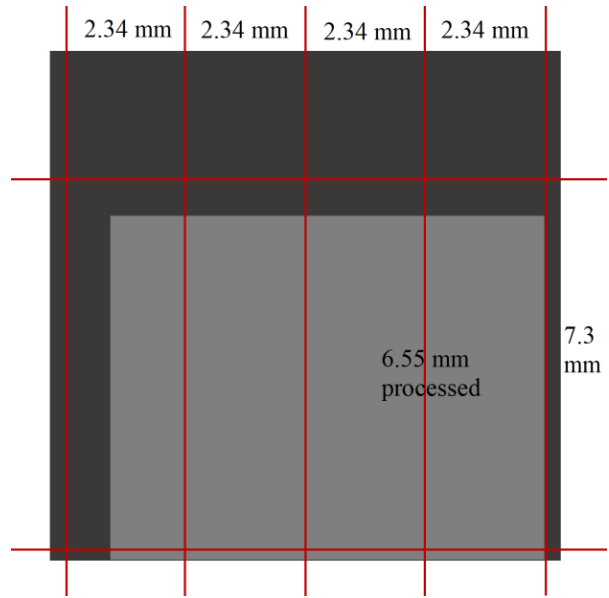

**Figure S1.** Schematic of diced laser processed samples with the size of 7.30 mm × 2.34 mm.

**Supporting Information 2: Chemical Reactions of Etching Solutions**

Subsurface laser modified Si regions are intended to be selectively etched by $Cu(NO_3)_2$-based wet-chemical etching solution. The solution is a mixture of $Cu(NO_3)_2$, $HNO_3$, HF, $CH_3COOH$, and DI water. The etching mechanism and expected reactions are as follows:

Equation S 1 is the reduction of Cu ions at the surface of the cathode. In the mechanism, Si injects electron to adsorbed Cu. Equation S 2 is the reduction of $HNO_3$ at the surface of the cathode which is faster at the presence of Cu nanoparticles. When $HNO_3$ is added to the solution, hole generation rate drastically increases and it becomes higher than the hole consumption during the etching process. The Cu ions are reduced to solid Cu by attracting minority negative charge carriers available from Si in Equation S 1 and Equation S 2 are used either in the direct dissolution reaction of Si at anode (Equation S 3) or in oxidation (Equation S 4) followed by etching (Equation S 5) of the formed ultra-thin $SiO_2$ on Si surface by HF [1]–[3].

$$Cu^{2+} + 2e_{VB}^- \rightarrow Cu^0(s) \qquad\qquad E^0 = +0.34 \text{ V/NHE} \quad \textbf{(S 1)}$$

$$HNO_3 + 3H^+ \rightarrow NO + 2H_2O + 3h^+ \qquad\qquad E^0 = +0.96 \text{ V/NHE} \quad \textbf{(S 2)}$$

$$Si + 4HF_2^- \rightarrow SiF_6^{2-} + 2HF + H_2 + 2e^- \qquad\qquad E^0 = -1.20 \text{ V/NHE} \quad \textbf{(S 3)}$$

$$Si + 2H_2O \rightarrow SiO_2 + 4H^+ + 4e^- \qquad\qquad E^0 = -0.84 \text{ V/NHE} \quad \textbf{(S 4)}$$

$$SiO_2 + 6HF \rightarrow H_2SiF_6 + 2H_2O \qquad\qquad E^0 = +0.33 \text{V/NHE} \quad \textbf{(S 5)}$$

Both $Cu^{2+}$ and $HNO_3$ act as oxidants while Si serves as the reductant in abovementioned redox system. Although Si etching rate in $HNO_3$/HF solution is low in the absence of metallic Cu ions, etching is accelerated by adding few moles of $Cu(NO_3)_2$ as source of Cu ions.

In order to selectively etch laser processed regions with minimal damage to unprocessed c-Si regions, the concentration of each chemical constituents of the solution is varied as shown in Table 1. in main manuscript.

Wet etching process of Si is feasible since both Cu and Si have almost similar electronegativities (1.9 eV of Cu and 1.8 eV of Si) allowing $Cu^{+2}$ ions to obtain electrons from Si atoms. In this way, Si is oxidized to $SiO_2$ and etched by HF. In addition, $HNO_3$ oxidizes the Si surface forming $SiO_2$ which is etched by the presence of HF. In the solution, $CH_3COOH$ acts as a diluent and reduces the overall reactant concentration.

[1]     C. Chartier, S. Bastide, and C. Lévy-Clément, "Metal-assisted chemical etching of silicon in HF-H2O2," *Electrochim. Acta*, vol. 53, no. 17, pp. 5509–5516, 2008, doi: 10.1016/j.electacta.2008.03.009.

[2]     Y. Wang, Y. Liu, L. Yang, W. Chen, X. Du, and A. Kuznetsov, "Micro-structured inverted pyramid texturization of Si inspired by self-assembled Cu nanoparticles," *Nanoscale*, vol. 9, no. 2, pp. 907–914, 2017, doi: 10.1039/c6nr08126f.

[3]     W. Chen *et al.*, "Difference in anisotropic etching characteristics of alkaline and copper based acid solutions for single-crystalline Si," *Sci. Rep.*, vol. 8, no. 1, pp. 2–9, 2018, doi: 10.1038/s41598-018-21877-x.

**Supporting Information 3: Comparison of different Chemical etchant composition**

In the first attempts, the effects of different etching solution compositions were investigated to find the most suitable solution for the subsurface laser modified region which can provide high selectivity at reasonable etch rate of modified regions. Table S1 shows the chemical composition of the considered etching solutions. The solution with potassium hydroxide (KOH); as one of the traditional chemicals to etch Si, was heated to 80 °C prior the etching to activate the reaction. All other etching processes were performed at room temperature.

**Table S1.** Chemical composition of the different etching solutions. The KOH-assisted etching was done at 80°C while other treatments were performed at room temperature. The duration of etching was varied based on the etch speed of the corresponding etching solution.

| Solution | HF (39%) ml | HNO$_3$ (65%) ml | CH$_3$COOH ml | Cu(NO$_3$)$_2$ g | AgNO$_3$ g | KOH w/vol% | Duration min |
|---|---|---|---|---|---|---|---|
| 1 | - | - | - | - | - | 30.0 | 10 |
| 2 | 9.0 | 68.2 | 22.8 | - | - | - | 120 |
| 3 | 40.0 | 20.0 | - | - | 11.00 | - | 10 |
| 4 | 36.0 | 25.0 | 18.0 | 1.00 | - | - | 150 |
| 5 | 65.0 | 35.0 | - | 2.00 | - | - | 3 |

For these experiments, subsurface laser modifications with parallel lines were applied on the Si wafers to identify the effects each etchant on distinct modified regions. The SEM images of etched part of samples are given in Figure S2. As it can be seen, the KOH solution (Figure S2 (a)) was not effective enough to selectively open the laser modified regions. In addition, the high etch rate difference of KOH solution on the different Si planes leads to the formation of undesired pyramid textures on the surface. On the other hand, the HF:HNO$_3$:CH$_3$COOH solution (Figure S2 (b)) was successful to reveal and open the laser modified region, however, the etch rate is found to be too low. Solution 3 with HF:HNO$_3$:AgNO$_3$ (Figure S2 (c)) is found to be moderately selective, the etch rate of laser modified regions is modest, and it left many residual particles on the surface. The initial composition of HF:HNO$_3$:CH$_3$COOH:Cu(NO$_3$)$_2$ solution (Figure S2 (d)) provided the required etch rate with better selectivity yet the obtained Si surface morphology features several holes which needs to be improved. The HF:HNO$_3$:Cu(NO$_3$)$_2$ solution (Figure S2 (e)) without acetic acid was too fast which etched large portion of sample in a very short time. In addition, it was not favorable since it could not provide the required selectivity and led to deteriorated surface morphology.

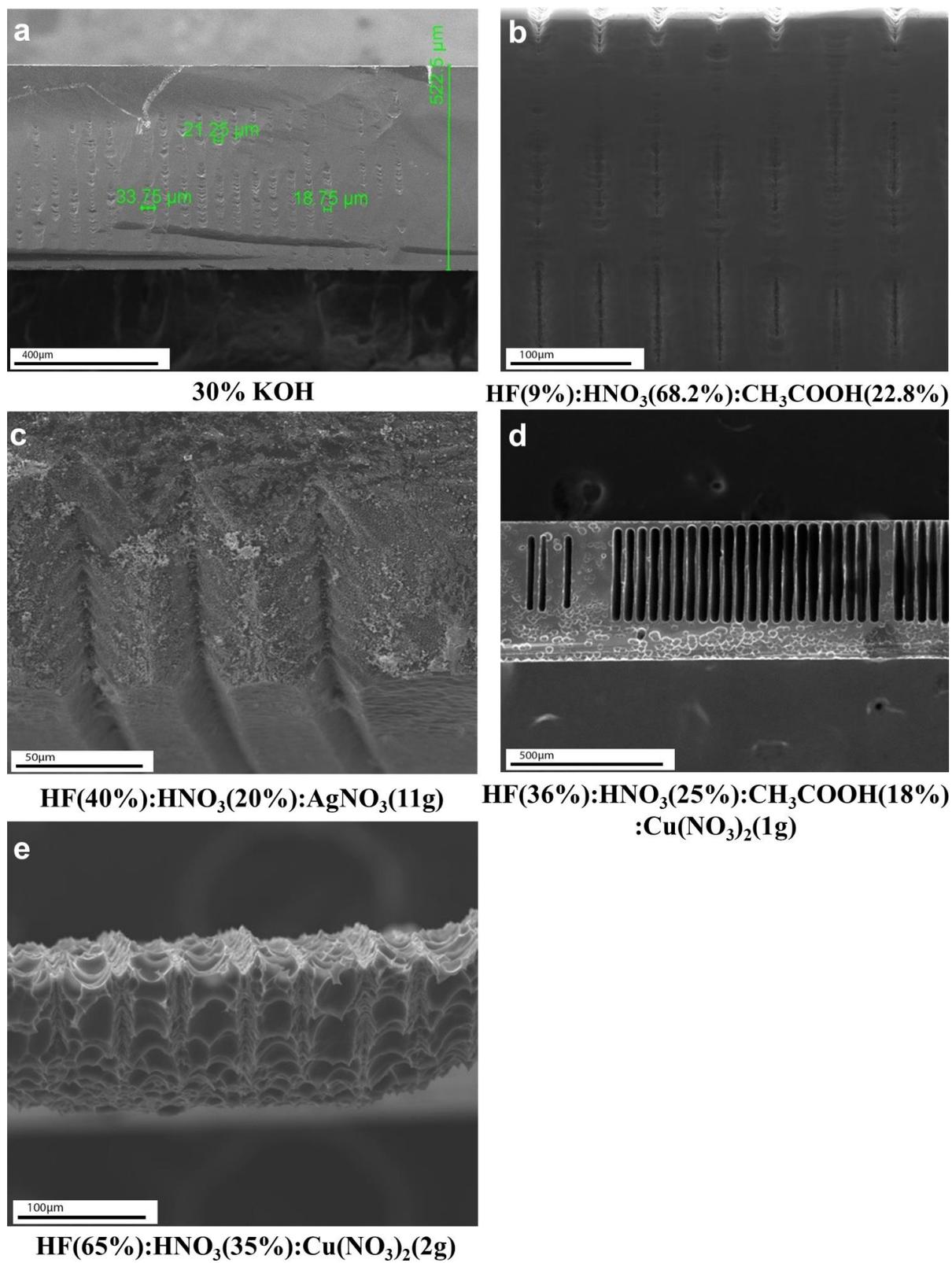

**30% KOH**

**HF(9%):HNO$_3$(68.2%):CH$_3$COOH(22.8%)**

**HF(40%):HNO$_3$(20%):AgNO$_3$(11g)**

**HF(36%):HNO$_3$(25%):CH$_3$COOH(18%)
:Cu(NO$_3$)$_2$(1g)**

**HF(65%):HNO$_3$(35%):Cu(NO$_3$)$_2$(2g)**

**Figure S2.** SEM images of different etching treatments on the subsurface laser-modified regions.
Subfigures (a) to (e) corresponds to solution 1 to 5 in the Table S1 respectively.

**Supporting Information 4: SEM images**

The SEM images of all sets categorized into 3 subsets representing variation of each etching component are presented in Figure S3, S4, and S5 respectively showing variation of $Cu(NO_3)_2$, HF, and $HNO_3$.

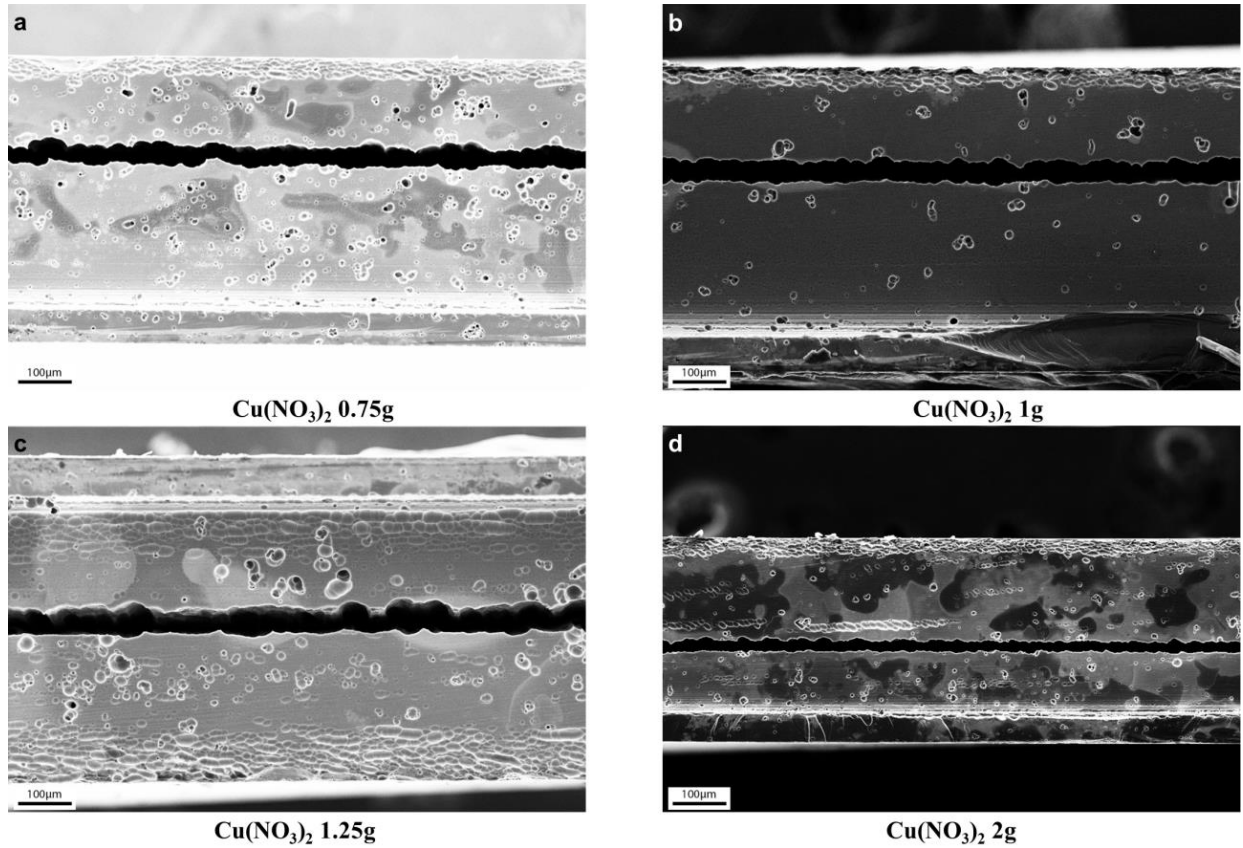

Cu(NO₃)₂ 0.75g          Cu(NO₃)₂ 1g

Cu(NO₃)₂ 1.25g          Cu(NO₃)₂ 2g

**Figure S3.** SEM images of laser modified openings after etching in the developed solution with different $Cu(NO_3)_2$ concentration: 0.75 g to 2.00 g in 100 ml of etching solution.

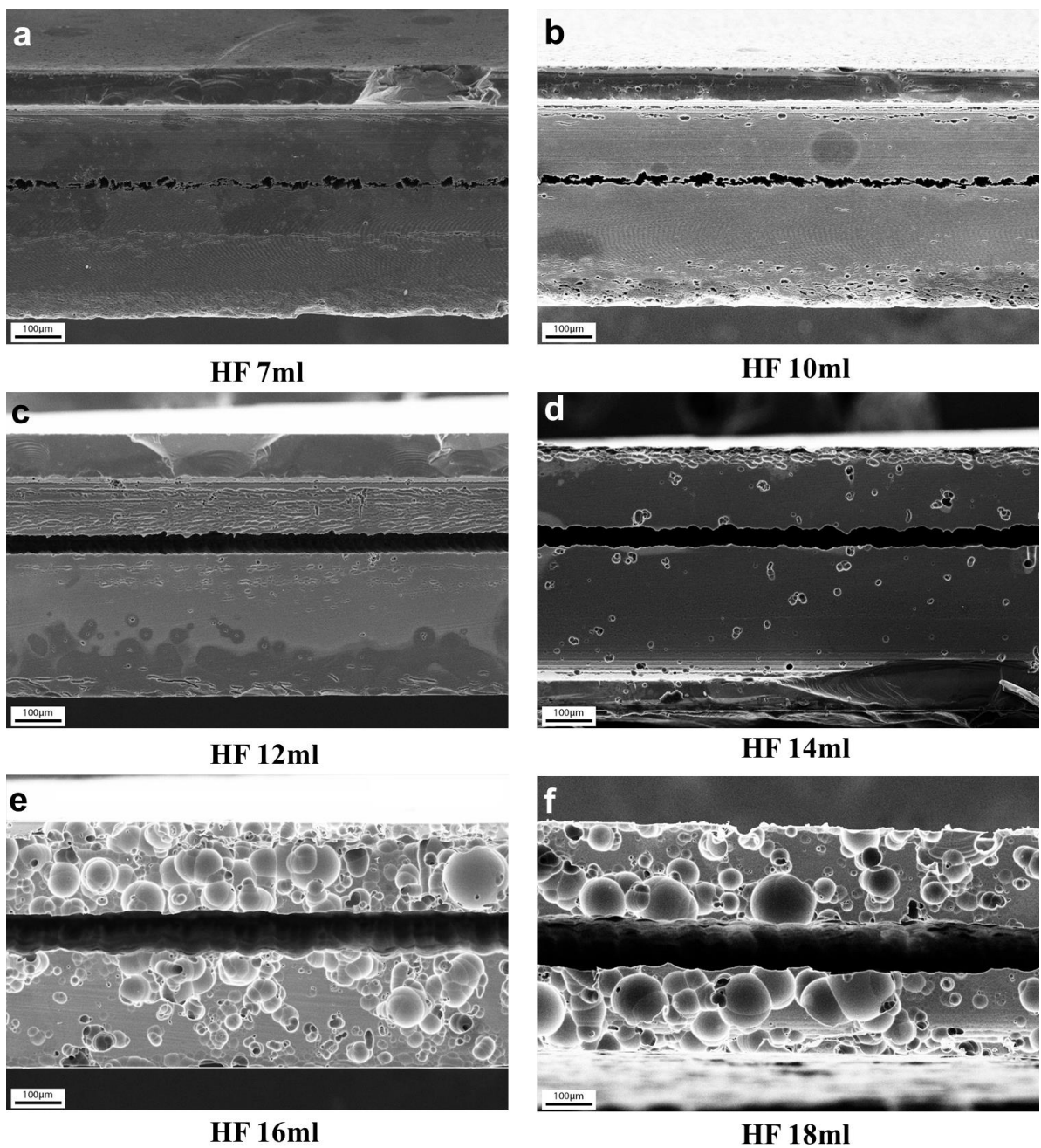

**Figure S4.** SEM images of variation of HF concentration from 31 ml to 46 ml in 100 ml of etching solution.

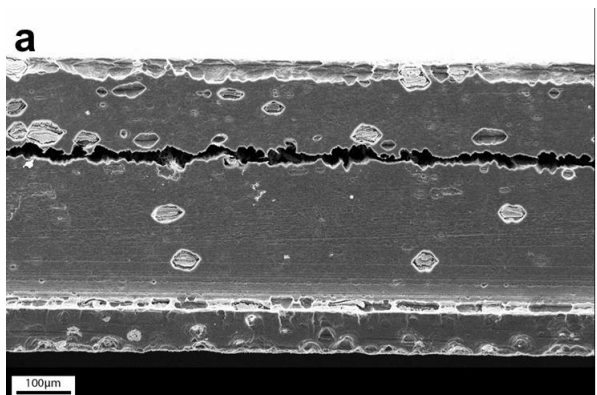

**HNO₃ 9.75ml**

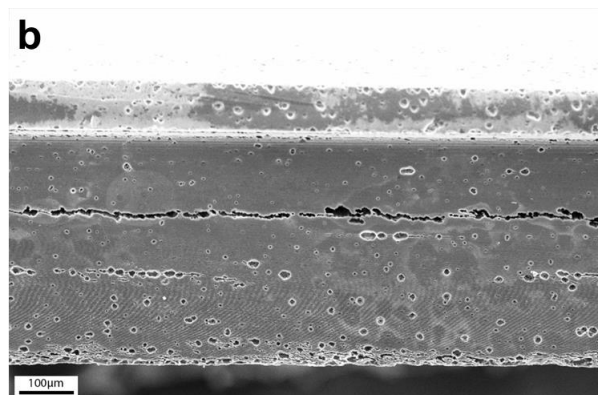

**HNO₃ 13ml**

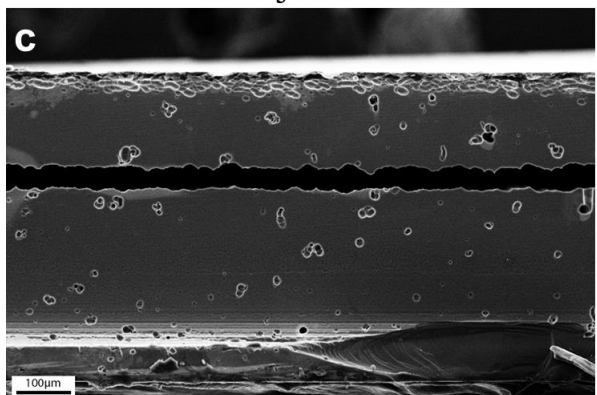

**HNO₃ 16.25ml**

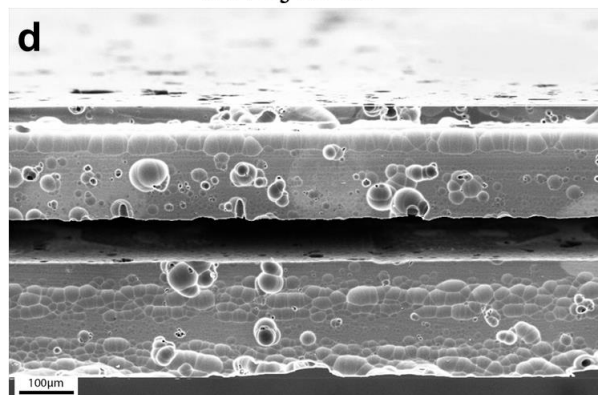

**HNO₃ 19.5ml**

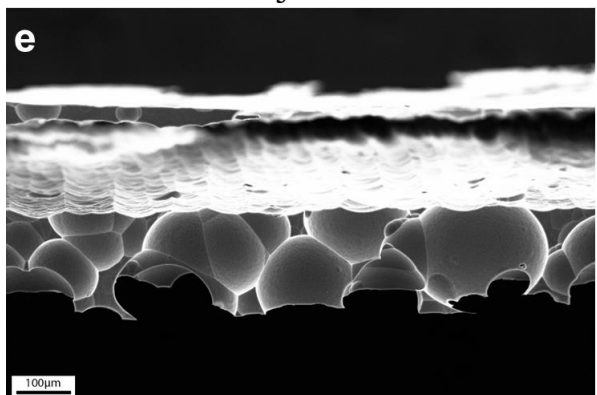

**HNO₃ 22.75ml**

**Figure S5.** SEM images of variation of HNO₃ concentration from 15 ml to 35 ml in 100 ml of etching solution.

**Supporting Information 5: Laser Modified Characterization**

The formation of surface structure induced by nanosecond-laser (ns-laser) processing is commonly followed by a series of phase transitions on the material's subsurface. In order to better analyze the structure of the laser-modified regions, high-resolution transmission electron microscopy (HRTEM) images were collected from double-side-polished <100> crystallographic Si sample processed by ns-laser from top surface and are shown in Figure S6 (a), (b). After laser processes, HRTEM revealed the formation of various grains (mixtures of amorphous and polycrystalline phases) with different orders and sizes. The selected area electron diffraction (SAED) patterns were presented in Figure S6 (c), (d). This allows to characterize the lattice dynamics of the modified regions. Nano structural developments include the formation of hollow concentric patterns and blurred concentric halo patterns; both dependent upon the initial nanostructure and the mechanism of the laser processing. Nano structured Si domains consist of a two-phase mixture of amorphous and locally ordered polycrystalline phases. The high-angle annular darkfield (HAADF) images of highly modified region (Figure S6 (e), (f)) further confirm the formation of both amorphous and locally ordered domain polycrystalline domains. In Figure S6, characteristics of amorphous phase are shown in (a), (c), and (e) and polycrystalline phase are shown in (b), (d), and (f) which are consistent through the specific atomic structure of the nanocrystalline modifications embedded in the Si by HRTEM and HAADF.

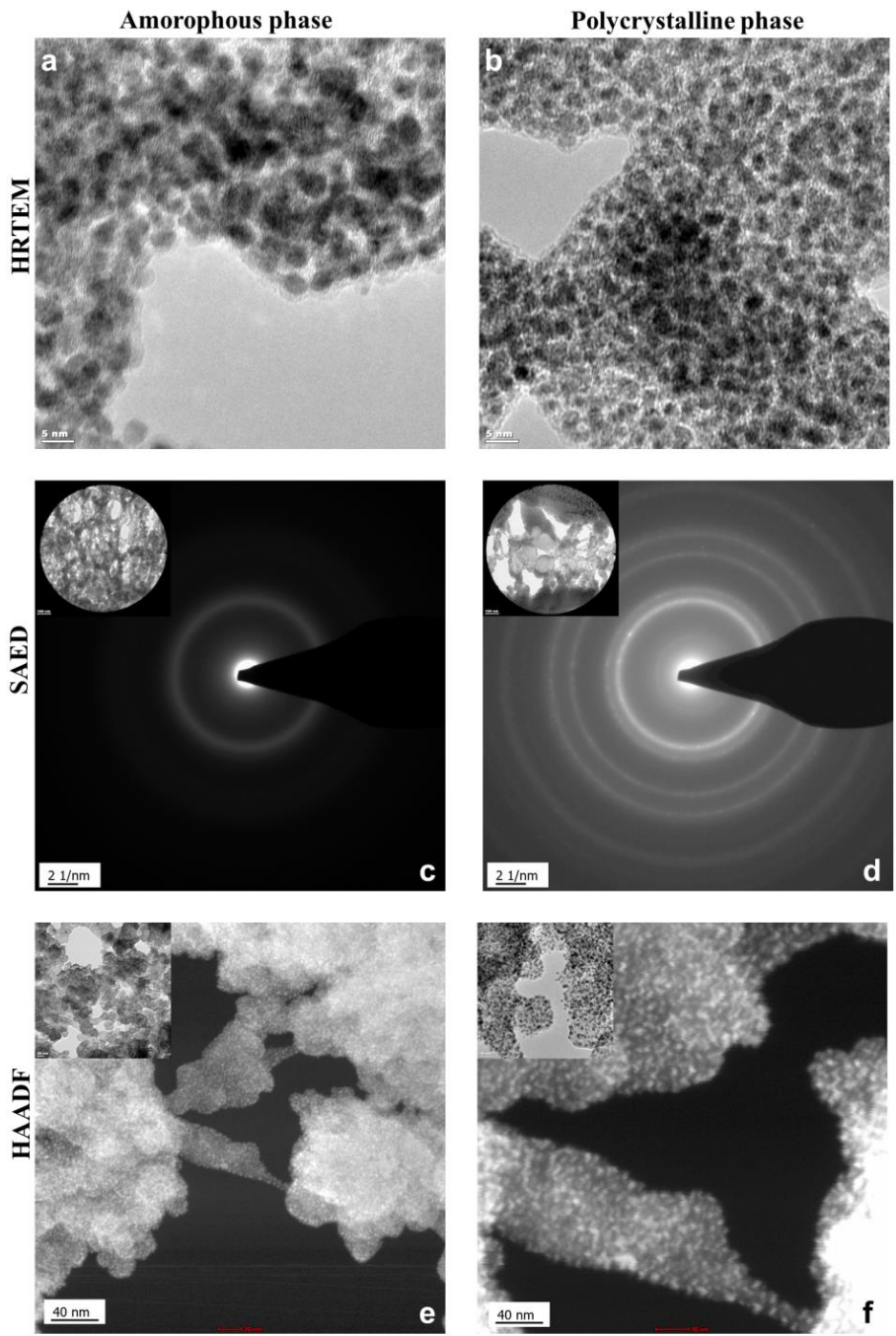

**Figure S6.** (a), (b) HRTEM images of the ordered domains of the laser modified regions. (c), (d) SAED patterns of laser modified regions with bright field TEM images of sample as inset. (e), (f) HAADF patterns of laser modified regions with bright field TEM images of sample as inset. (a), (c), (e) are amorphous phase and (b), (d), (f) are polycrystalline phase.

**Supporting Information 6: Correction Factor Calculations**

As discussed in *Result and Discussion* section, laser modified mass calculation requires three correction factors, namely, Correction Factor of Initial Mass (CFIM), Correction Factor of Etch Opening (CFEO), and Correction Factor of Attacked Area (CFAA).

We tend to match the shape and size of laser processed and unprocessed samples. Thus, we used the following CFIM to compensate the small discrepancies in the masses of the processed and unprocessed samples, defined as:

$$CFIM = \frac{Initial\ Mass\ of\ Unprocessed\ Sample}{Initial\ Mass\ of\ Laser\ Processed\ Sample} \tag{S 6}$$

Next, CFEO was determined by analyzing the opening of the etched part from the SEM images (a representative opening for *the champion etchant* with composition of $HF:HNO_3:CH_3COOH:H_2O$ – 56:65:72:207 with 0.01 g/ml $Cu(NO_3)_2.3H_2O$ is shown in Figure S7).

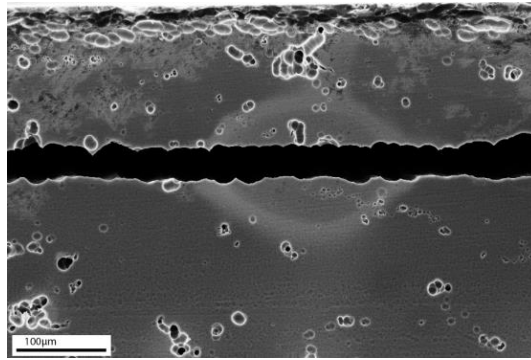

**Figure S7.** Cross-sectional SEM image after 60 min etching showing the opening and etch pits resulted from the etching process for the champion etchant.

This correction factor is essential for the samples where the openings after chemical etching are larger than the laser modified initial thickness (30 µm). In this case, all laser-modified regions are etched until a certain depth, exposing some parts of unprocessed regions above and below the laser modified region to the etchant. We assume that the etching of unprocessed regions is relatively proportional to the etching of unprocessed sample which experience exactly the same solution. On the other hand, for the samples with openings being less than the laser modified thickness, this correction factor is not used. To simplify the analysis, we assume that the geometrical factor is linear. The formulation of CFEO is given in Equation S 7.

$$CFEO = \frac{Corresponding\ Opening\ Thickness - 30\ \mu m}{Thickness\ Reduction\ of\ Unprocessed\ Sample} \tag{S 7}$$

The proportionality of etched part of unprocessed Si regions of laser processed sample to unprocessed sample is their relative difference in thickness. Since the openings are distinct straight lines, the "Area of Opening of Laser Processed Region of Laser Processed Sample" is calculated using Gwyddion [4] which is colored red in Figure S8. Then, this area is divided by the "Length of Opening at Given Magnification" which leads to "Corresponding Opening Thickness" as described in Equation S 7. As an example, the opening of laser processed region for *the champion etchant* is shown in Figure S8. In this figure, magnification is 500x which indicates that "Length of Opening at Given Magnification" is 596 µm.

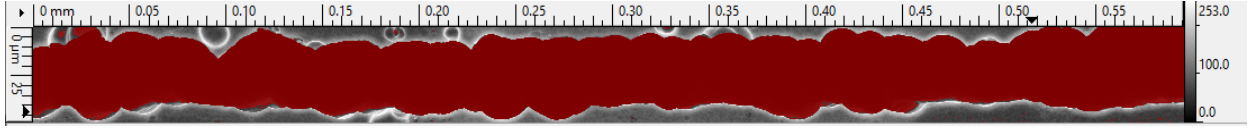

**Figure S8.** Area of the opening after etching treatment at the specific magnification (Set Opt: 500x)

$$Corresponding\ Opening\ Thickness =$$
$$\frac{Area\ of\ Opening\ of\ Processed\ Region\ of\ Laser\ Processed\ Sample}{Length\ of\ Opening\ at\ Given\ Magnification} \tag{S 8}$$

Based on the fact that the subsurface laser modified region thickness is 30 µm, the thickness of unprocessed region is "Corresponding Opening Thickness" minus 30 µm.

The "Thickness Reduction of Unprocessed Sample" is calculated based on the "Exact Unprocessed Thickness" which can be obtained from initial mass converted to volume using density as stated in Equation S 9.

$$Exact\ Unprocessed\ Thickness = \frac{Initial\ Mass\ of\ Laser\ Processed\ Sample}{Silicon\ Density} \times$$
$$\frac{1}{Diced\ Area\ (2.34mm\ \times 7.3mm)} \tag{S 9}$$

In order to calculate "Thickness Reduction of Unprocessed Sample", "Final Volume" given Equation S 10 is required. "Thickness Reduction of Unprocessed Sample" ($x$) can be obtained by considering the thickness reduction from all 3 dimensions and solving the Equation S 11.

$$Final\ Volume = \frac{Final\ Mass\ of\ Laser\ Processed\ Sample}{Silicon\ Density} \tag{S 10}$$

$$(7.3 - x)(2.34 - x)(Exact\ Laser\ Processed\ Thickness - x) - Final\ Volume \tag{S 11}$$

The mass reduction from unprocessed samples only occurs on the outer surfaces, being the c-Si lattice. However, in the case of laser-modified samples, etching occurs on both the outer surfaces as well as in the inner laser-modified region. It is worth mentioning that the volume neighboring the laser-modified region gets damaged during laser processing. The mass reduction of the outer layer of both laser-processed and unprocessed samples are just identical. Thus, by applying CFIM and CFEO for each set, the laser modified mass can be calculated as:

$$Laser\ Modified\ Mass = Mass\ Reduction\ of\ Laser\ Processed\ Sample -$$
$$\frac{Mass\ Reduction\ of\ Unprocessed\ Sample}{CFIM} - Mass\ Reduction\ of\ Unprocessed\ Sample \times\ CFEO \tag{S 12}$$

Later, for better comparison between the laser modified region of processed sample and the unprocessed sample, we consider the percentage of total mass by dividing each mass change by its own total mass:

$$\%\ Mass\ Change\ = \frac{Mass\ Change}{Total\ Mass} \tag{S 13}$$

Where the "Total Mass" of the laser-modified region is calculated based on the laser modified volume, being 6.55×2.34×0.03 mm$^3$ and the Si density being 2.329 mg/mm$^3$.

After obtaining the %Mass Change of laser modified region over time, we calculate the etch rates. For the champion etchant, the etch mass is shown in Figure S9 for both unprocessed sample and laser-modified region of the processed samples.

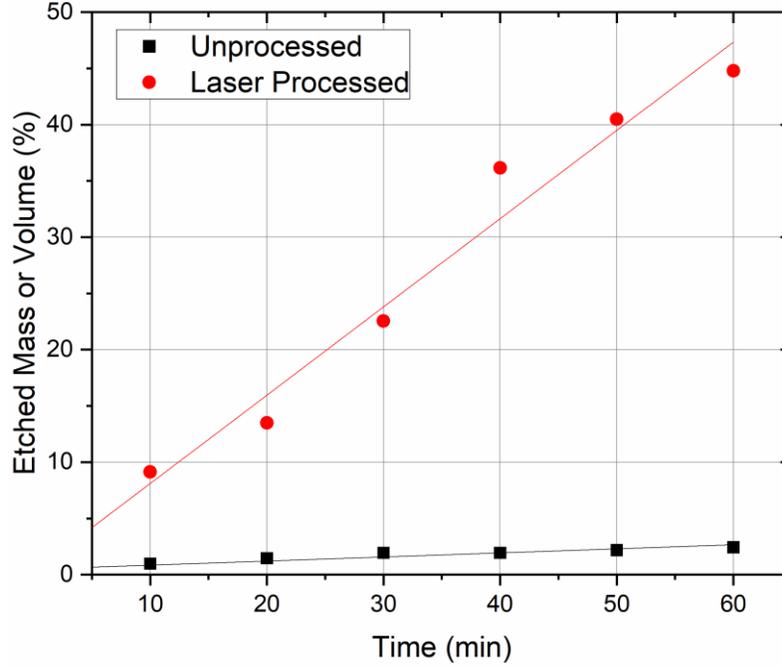

**Figure S9.** The etch rates of laser modified region (red) and unprocessed sample over time, including the linear trend line of each etch rate for the champion etchant.

The huge difference between the masses of the unprocessed sample and the laser modified regions; of the laser processed sample, is due to the exposed area of the sample and regions to the etchant. The unprocessed sample is exposed to the etchant from all surfaces ($7.4\times2.34\times0.525$ mm$^3$), whereas the laser-modified region is subjected to etchant from only 3 surfaces, as illustrated in yellow in Figure 1 (b) in the manuscript.

Since the laser-modified region of the laser processed sample and unprocessed sample are attacked from different surfaces with different areas, the selectivity needs to be corrected by this area factor. The CFAA is defined as the "Attacked Area of Unprocessed Sample" over the "Attacked Area of Laser Modified Region" (a constant here; Equation S 14).

$$CFAA = \frac{Attacked\ Area\ of\ Unpcessed\ Sample}{Attacked\ Area\ of\ Laser\ Modified\ Region} = \frac{2(a*b+a*c+b*c)}{2(a'*c')+b'*c'} = 95.609 \qquad \textbf{(S 14)}$$

where,

a = 7.4, b = 2.34, c = 0.525 and a' = 6.55, b' = 2.34, c' = 0.03

To calculate the selectivity, first, we regress the mass reduction of the laser-modified region and unprocessed sample to obtain the corresponding average etch rates. Next, the selectivity was defined as the etch rate of laser modified region divided by etch rate of unprocessed sample, in each set using:

$$Selectivity = \frac{Etch\ Rate\ of\ Laser\ Modified\ Region}{Etch\ Rate\ of\ Unprocessed\ Sample} \times CFAA \qquad \textbf{(S 15)}$$

For the SALP processing, after obtaining the %Mass Change of laser modified region over time, we calculate the etch rates. An example for the champion etchant is shown in Figure S10 for both unprocessed sample and laser modified region.

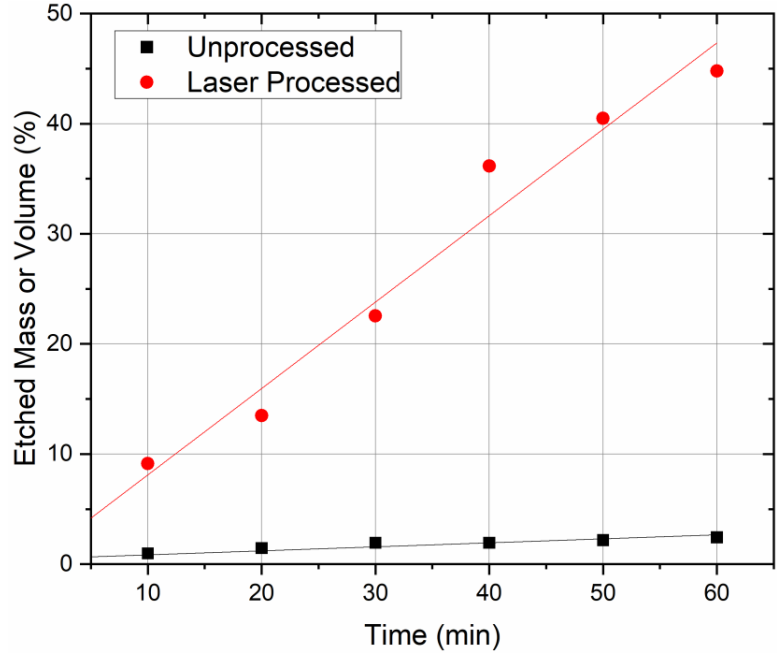

**Figure S10.** Etched masses of laser modified region and unprocessed sample corrected by exposed area for the champion etchant.

For the SALP structuring, the opening area of the laser modified region ($2\times2=4$ mm$^2$) is 9 times larger than that of CALP ($6.55\times0.03+6.55\times0.03+2.34\times0.03=0.46$ mm$^2$). The dramatic difference between the masses of the unprocessed sample and the laser modified regions is due to the exposed area of the sample and regions to the etchant. The unprocessed sample is exposed to the etchant from all surfaces ($10\times10\times1$ mm$^3$). In contrast, the laser modified region is subjected to etchant from the top on SALP, as illustrated in yellow in Figure 1 (d).

Since in the SALP structuring, we eliminated the limitation of diffusion in the etching region, the mass change of laser modified region can be directly calculated by subtracting the etched mass of the laser modified sample from unprocessed sample which is normalized by total mass reduction of laser processed sample (Equation S 16); to include the total change of laser processed sample as shown in Figure 2 (b).

$$Laser\ Modified\ Mass\ Change = \frac{Mass\ Reduction\ of\ Laser\ Processed\ Sample - Mass\ Reduction\ of\ Unprocessed\ Sample}{Total\ Mass\ reduction\ of\ laser\ Processed\ Sample}\ \text{(S 16)}$$

The ratio of laser modified region to unprocessed sample is presented by Equation S 17, indicating that what portion of reduction is related to which region at each time.

$$Ratio\ of\ Laser\ Modified\ Mass\ to\ Unmodified = \frac{Mass\ Reduction\ of\ Laser\ modified\ Region}{Mass\ reduction\ of\ Unprocessed\ Sample}\text{(S 17)}$$

[4]     D. Nečas and P. Klapetek, "Gwyddion: an open-source software for SPM data analysis," *Open Phys.*, vol. 10, no. 1, Jan. 2012, doi: 10.2478/s11534-011-0096-2.

**Supporting Information 7: IR Camera Images**

We measured the cross section of etched area of the laser modified region from the images obtained by the infrared (IR) camera. Figure S11 (a) clearly shows the laser modified and etched regions. The IR image of *the Champion Etchant* after 60 min etching is shown in Figure S11 (b). In this image, the black color is the etched modified region. In addition, the grey color at the center is laser modified region while the grey color at top of image is unmodified region. The two grey regions look similar since the modifications in crystallinity is not profound to be visible in IR camera region. The opening is obtained using SEM images in Figure S3, S4, and S5.

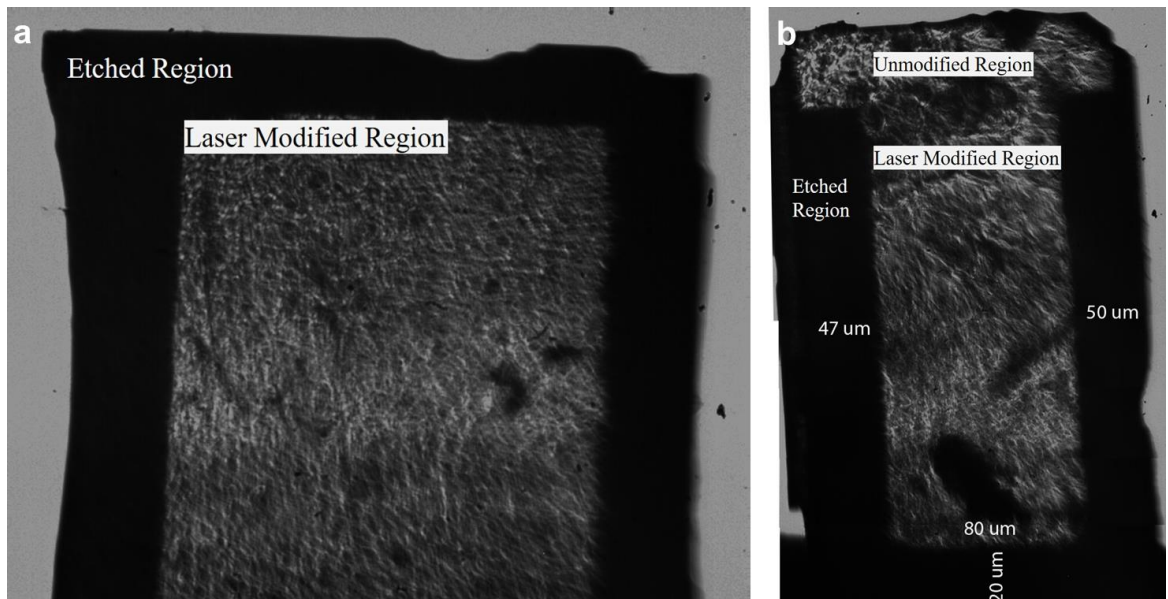

**Figure S11.** Image of infrared camera of the etched laser modified region.

**Supporting Information 8: Further analysis on the effects of solution components on SALP and CALP**

Comparison of mass changes between CALP and SALP is shown in Figure S12 (a), (c), (e), indicating the effect of subsurface and surface etching. The ratio of area normalized etched mass of the SALP samples are plotted in Figure S12 (b), (d), (f).

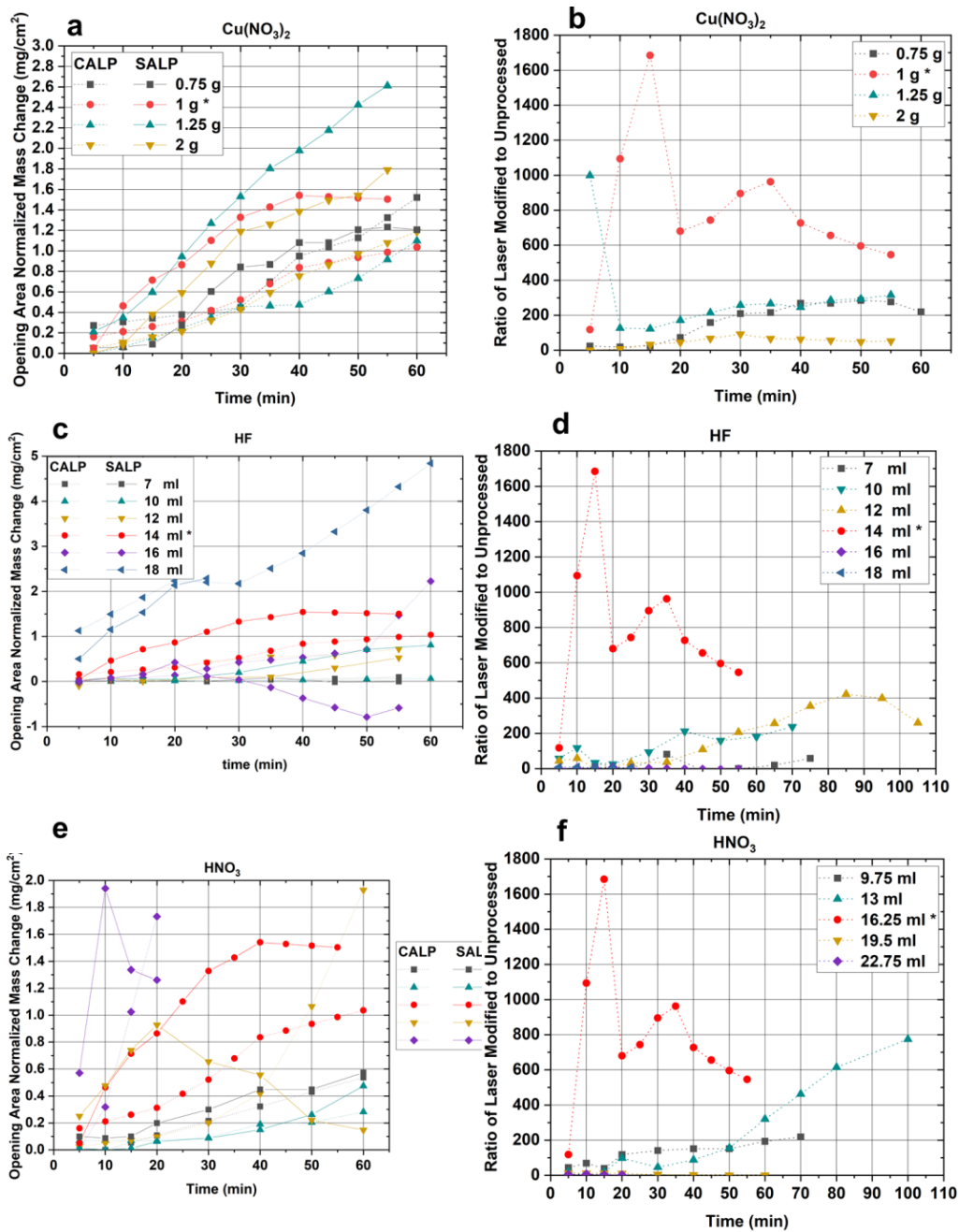

**Figure S12.** (a), (c), (e) Area corrected etched mass of laser modified region for SALP (solid line) and CALP (dashed line). (b), (d), (f) Ratio of area normalized etched mass of laser modified region over the unprocessed region for SALP. The changing parameters are (a), (b) Cu(NO3)2

concentrations from 0.75 to 2.00 g, (c), (d) HF concentrations from 7 ml to 18 ml, (e), (f) HNO3 concentrations from 9.75 to 22.75 ml in 100 ml of solution.

In Figure S12 (a), the etched masses are normalized to their opening area due to the difference in the opening area exposed to etchant solution. The laser-modified etched mass in SALP exhibits a faster etching compared to that of the subsurface laser processing. In Figure S12 (b), the ratio of laser modified to unprocessed samples decreases as etching time increases. For longer durations, the etchant starts etching unmodified parts when all the modified masses are etched away; thus, the ratio begins to increase again, and the selectivity will therefore decrease. In Figure S12 (c), the laser modified etched mass in SALP goes higher than subsurface except for the solution with higher HF concentrations. In Figure S12 (d), the ratio of laser modified region to unprocessed sample decreases as time increases for the lower HF concentration. In Figure S12 (e), the laser modified etched mass in SALP goes higher than subsurface except for a higher dose concentration of $HNO_3$.

**Supporting Information 9: Comparison of MEMC and the Champion Etchant**

We adopted the MEMC chemical compositions and its initial concentrations since $Cu(NO_3)_2$ is very defect sensitive which is effective not only on c-Si but also on laser modified Si regions. Even though it is orientation and dopant type dependent, it can be used for both dopant types since it has reasonable etching durations for different purposes and long lifetime without spiking in solution. It also does not require excessive heating and not forming dangerous gasses in solution. Additionally, unlike other etching solutions that contain Cr (VI) as an oxidizing agent such as Sirtl, Seiter, Wright, and Secco, MEMC etching is Cr-free. However, it is important to consider that the conventional MEMC solution is too fast in such a way that it removes more than 70% of our wafer (<100> p-type crystalline Si wafer) in just 12 min. Considering MEMC low selectivity, the wafer fully dissolved in 20 min.. Thus, we modified and optimized the MEMC solution to obtain the optimum solution (*the Champion Etchant*). In *the Champion Etchant*, the etch rate is lower which provides required time for chemical treatments depending on the desired application. It is also more dopant type dependent and more anisotropic. After considering different concentrations of each component in the solution, the optimum concentration is achieved (14 (HF):16.25 ($HNO_3$):18 ($CH_3COOH$):51.75 DI water with 1g $Cu(NO_3)_2.3H_2O$).

The comparisons between MEMC and *the Champion Etchant* in terms of orientation and dopant type dependencies on crystalline Si wafers are presented in Table S2 and Table S3.

**Table S2.** Effect of MEMC etching solution on the <111> and <100> plane and p-type and n-type of crystalline Si wafers with duration of 12 min.

|   | Type | Plane | Thickness μm | Initial Mass mg | Final Mass mg | Mass Reduction mg | Normalized Mass Reduction % |
|---|------|-------|-----------|--------------|------------|----------------|------------------------|
| 1 | p-type | 111 | 525±25 | 274.55 | 139.95 | 134.60 | 49.03 |
| 2 | p-type | 100 | 279±15 | 150.85 | 044.65 | 106.20 | 70.40 |
| 3 | n-type | 100 | 279±15 | 144.05 | 092.65 | 051.40 | 35.68 |

**Table S3.** Effect of *the Champion Etchant* etching solution on the <111> and <100> plane and p-type and n-type of crystalline Si wafers with duration of 18 min.

|   | Type | Plane | Thickness μm | Initial Mass mg | Final Mass mg | Mass Reduction mg | Normalized Mass Reduction % |
|---|------|-------|-----------|--------------|------------|----------------|------------------------|
| 1 | p-type | 111 | 525±25 | 275.95 | 275.10 | 000.85 | 0.31 |
| 2 | p-type | 100 | 279±15 | 149.65 | 148.50 | 001.15 | 0.77 |
| 3 | n-type | 100 | 279±15 | 144.20 | 142.50 | 001.70 | 1.18 |

According to Table S2 and Table S3, it can be recognized *the Champion Etchant* etches p-type 1.48 times slower than n-type while MEMC etches p-type 2.07 times faster than n-type. In addition, <100> plane etch rate is higher than <111> for both MEMC and *the Champion Etchant*.

**Supporting Information 10: XRD (Crystallinity) Analysis**

To understand the nature of laser modified regions, EBSD, XRD, and Raman analyses were performed. These different techniques confirm that the Si regions experienced laser process exhibit a rather nanocrystalline phase. During laser process, the Si melts, and recrystallizes leaving nanocrystalline separated by grain boundaries which appears in the EBSD analysis (Figure S13). The broadening in full width half maximum (FWHM) in Raman spectra for the Si after laser process (Figure S14), demonstrates the formation of nanocrystalline phase. Additionally, XRD analysis shows the formation of different diffraction peaks for Si after laser process (Figure S15). These peaks disappear after 30 minutes etching. The peak that appears after etching at ~52° is likely associated to $SiN_x$ [5]. Since the used etchant contains $HNO_3$ and $Cu(NO_3)_2$, it seems that there are Si-N-terminated portions on the surface. The peak at ~52° also appears for the sample after laser process, which can be attributed to the decomposition of N from the air during the laser process forming Si-N bonds on the surface.

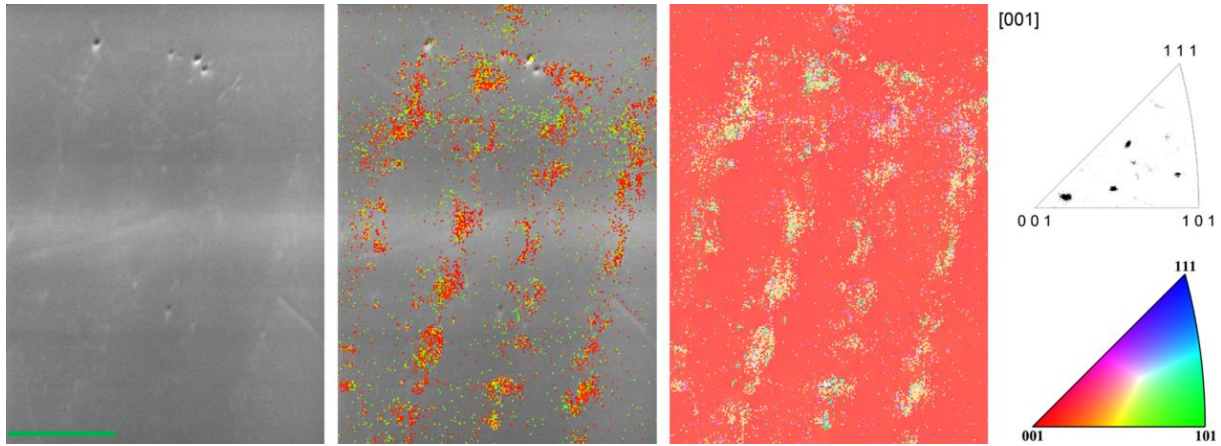

**Figure S13.** EBSD analysis of laser modified Si.

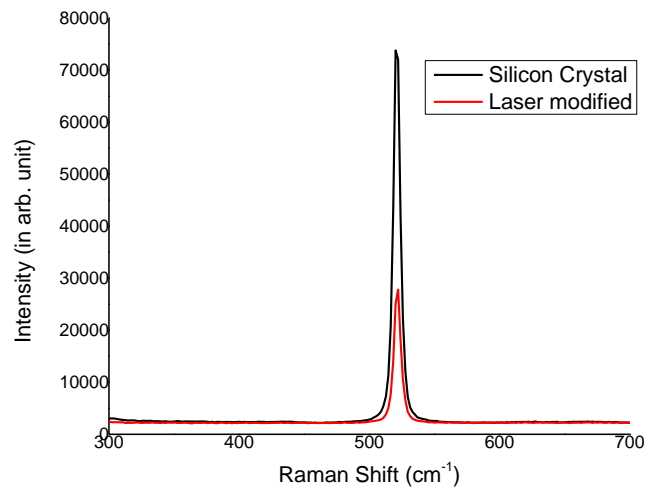

**Figure S14.** Raman analysis of laser modified Si and crystalline silicon.

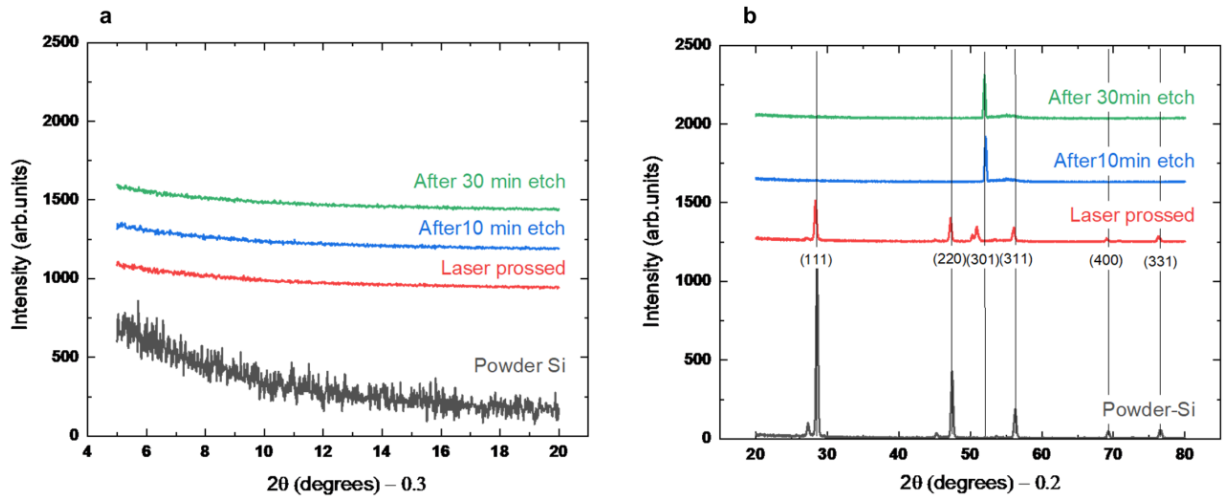

**Figure S15.** XRD spectra of Si after different processes steps: laser processed, after 10 min etching, and after 30 min etching. Additionally, the XRD spectrum of Si powder is given as reference. Grazing incidence (a) from 5° to 20° (b) from 20° to 80°.

[5] F. Tiour *et al.*, "Opto-structural properties of Si-rich SiNx with different stoichiometry," *Appl. Phys. A Mater. Sci. Process.*, vol. 126, no. 1, pp. 1–10, 2020, doi: 10.1007/s00339-019-3258-5.